\documentclass[12pt,letterpaper]{article}
\usepackage{changepage}
\usepackage[utf8]{inputenc}
\usepackage{textcomp,marvosym}
\usepackage{fixltx2e}
\usepackage{amsmath,amssymb}
\usepackage{cite}
\usepackage{nameref,hyperref}
\usepackage[right]{lineno}
\usepackage{microtype}
\DisableLigatures[f]{encoding = *, family = * }
\usepackage{rotating}
\usepackage[aboveskip=1pt,labelfont=bf,labelsep=period,justification=justified,singlelinecheck=off]{caption}
\makeatletter
\renewcommand{\@biblabel}[1]{\quad#1.}
\makeatother
\usepackage{amsthm,amsfonts}
\usepackage[english]{babel}
\usepackage{epstopdf}
\usepackage{epsfig}
\date{}

\newcommand{\NS}{NS}
\newcommand{\STD}{STD}
\newcommand{\SFA}{SFA}

\begin{document}
\vspace*{0.35in}

\begin{flushleft}
{\Large
\textbf\newline{Network events on multiple space and time scales in cultured neural networks and in a stochastic rate model}
}
\newline
\\
Guido Gigante\textsuperscript{1,2,*},
Gustavo Deco\textsuperscript{3,4},
Shimon Marom\textsuperscript{5},
Paolo Del Giudice\textsuperscript{1,6}
\\
\bf{1} Italian Institute of Health, viale Regina Elena 199, Rome, 00161, Italy
\\
\bf{2} Mperience srl, via Elea 8, 00183, Rome, Italy
\\
\bf{3} Center for Brain and Cognition, Computational Neuroscience Group, Department of Information and Communication Technologies, Universitat Pompeu Fabra, Roc Boronat 138, Barcelona, 08018, Spain
\\
\bf{4} Institució Catalana de la Recerca i Estudis Avançats (ICREA), Universitat Pompeu Fabra, Passeig Lluís Companys 23, Barcelona, 08010, Spain
\\
\bf{5} Technion - Israel Institute of Technology, Haifa 32000, Israel
\\
\bf{6} INFN, Sezione di Roma 1, Italy
\\

* E-mail: Corresponding guido.gigante@mperience.com
\end{flushleft}

\section*{Abstract}
Cortical networks, \textit{in-vitro} as well as \textit{in-vivo}, can spontaneously generate a variety of collective dynamical events such as network spikes, UP and DOWN states, global oscillations, and avalanches. Though each of them have been variously recognized in previous works as expressions of the excitability of the cortical tissue and the associated nonlinear dynamics, a unified picture of their determinant factors (dynamical and architectural) is desirable and not yet available. Progress has also been partially hindered by the use of a variety of statistical measures to define the network events of interest.
We propose here a common probabilistic definition of network events that, applied to the firing activity of cultured neural networks, highlights the co-occurrence of network spikes, power-law distributed avalanches, and exponentially distributed `quasi-orbits', which offer a third type of collective behavior.
A rate model, including synaptic excitation and inhibition with no imposed topology, synaptic short-term depression, and finite-size noise, accounts for all these different, coexisting phenomena. We find that their emergence is largely regulated by the proximity to an oscillatory instability of the dynamics, where the non-linear excitable behavior leads to a self-amplification of activity fluctuations over a wide range of scales in space and time. In this sense, the cultured network dynamics is compatible with an excitation-inhibition balance corresponding to a slightly sub-critical regime.
Finally, we propose and test a method to infer the characteristic time of the fatigue process, from the observed time course of the network's firing rate. Unlike the model, possessing a single fatigue mechanism, the cultured network appears to show multiple time scales, signalling the possible coexistence of different fatigue mechanisms.

\section*{Introduction}
The spontaneous activity of excitable neuronal networks exhibits a spectrum of dynamic regimes ranging from quasi-linear, small fluctuations close to stationary activity, to dramatic events such as abrupt and transient synchronization. Understanding the underpinnings of such dynamic versatility is important, as different spontaneous modes may imply in general different state-dependent response properties to incoming stimuli and different information processing routes.

Cultured neuronal networks offer a controllable experimental setting to open a window into the network excitability and its dynamics, and have been used intensively for the purpose.

Neuronal cultures in early development phases naturally show alternating quasi-quiescent states and `network spikes' (\NS{}) of brief outbreaks of network activity \cite{Baltz2011,eytan2006dynamics, giugliano2004single,gritsun2011experimental,Park2006a, Wagenaar2006}.

In addition, recent observations \textit{in-vitro} (and later even \textit{in-vivo}) revealed a rich structure of network events (`avalanches') that attracted much attention because their spatial and temporal structure exhibited features (power-law distributions) reminiscent of what is observed in a `critical state' of a physical system (see e.g. \cite{beggs2003neuronal,petermann2009spontaneous}, and \cite{plenz2007organizing,plenz2014criticality} and references therein). Generically, an avalanche is a cascade of neural activities clustered in time; while there persist ongoing debate on the relation between observed avalanches and whatever `criticality' may mean for brain dynamics \cite{linaro2011inferring}, understanding their dynamical origin remains on the agenda.

Quasi-synchronous \NS{}, avalanches and small activity fluctuations are frequently coexisting elements of the network dynamics. Besides, as we will describe in the following, in certain conditions one can recognize network events which are clearly distinct from the mentioned network events, which we name here as `quasi-orbits'.

The abundant modeling literature on the above dynamical phenomena has been frequently focused on specific aspects of one of them \cite{millman2010self,levina2007dynamical}; on the other hand, getting a unified picture is made often difficult by different assumptions on the network's structure and constitutive elements and, importantly, by different methods used to detect the dynamic events of interest.

In the present paper we define a common probabilistic criterion to detect various coexisting dynamic events (\NS{}, avalanches and quasi-orbits) and adopt it to analyze the spontaneous activity recorded from both cultured networks, and a computational rate model. 

Most theoretical models accounting for \NS{} are based on an interplay between network self-excitation on one side, and on the other side some \textit{fatigue} mechanism provoking the extinction of the network spike \cite{millman2010self,levina2007dynamical}. For such a mechanism two main options, up to details, have been considered: neural `spike-frequency adaptation' \cite{giugliano2004single,Thivierge2008} and synaptic  `short-term depression' (\STD{}) \cite{gritsun2011experimental,Park2006a,Tsodyks2000,Vladimirski2008,benita2012synaptic,reig2006impact}. Despite their differences, both mechanisms share the basic property of generating an activity-dependent self-inhibition in response to the upsurge of activity upon the generation of a \NS{}, the more vigorously, the stronger the \NS{} (i.e. the higher the average firing rate). In this paper, we will mainly focus on \STD{}, stressing the similarities of the two mechanisms, yet not denying their possibly different dynamic implications.

While \STD{} acts as an activity-dependent self-inhibition, the self-excitability of the network depends on the balance between synaptic excitation and inhibition; investigating how such balance, experimentally modifiable through pharmacology, influences the dynamics of spontaneous \NS{}s is interesting and relevant as a step towards the identification of the `excitability working point' in the experimental preparation.

To study the factors governing the co-occurrence of different network events and their properties we adopt a rate model for the dynamics of the global network activity, that takes into accounts finite-size fluctuations and the synaptic interplay between one excitatory and one inhibitory population, with excitatory synapses being subject to \STD{}.

On purpose we implicitly exclude any spatial topology in the model, which is meant to describe the dynamics of a randomly connected, sparse network, since we intend to expose the exquisite implications of the balance between synaptic excitation and inhibition, and the activity-dependent self-inhibition due to \STD{}. In doing this, we purposely leave out not only the known relevance of a topological organization \cite{beggs2004neuronal,plenz2007organizing,massobrio2014criticality}, but also the role of cliques of neurons which have been proposed to play a pivotal role in the the generation of \NS{} as functional hubs \cite{luccioli2014clique}, as well as the putative role of `leader neurons'.

We perform a systematic numerical and analytical study of \NS{}s for varying excitation/inhibition balance. The distance from an oscillatory instability of the mean-field dynamics (in terms of the dominant eigenvalue of the linearized dynamics) largely appears to be the sole variable governing the statistics of the inter-\NS{} intervals, ranging from a very sparse, irregular bursting (coefficient of variation $\mathrm{CV} \sim 1$) to a sustained, periodic one ($\mathrm{CV} \sim 0$). The intermediate, weakly synchronized regime ($\mathrm{CV} \sim 0.5$), in which the experimental cultures are often observed to operate, is found in a neighborhood of the instability that shrinks as the endogenous fluctuations in the network activity become smaller with increasing network size.

Moreover, the model robustly shows the co-presence of avalanches with \NS{} and quasi-orbits. The avalanche sizes are distributed according to a power-law over a wide region of the excitation-inhibition plane, although the crossing of the instability line is signaled by a bump in the large-size tail of the distribution; we compare such distributions and their modulation (as well as the distributions of \NS{}) across the instability line with the experimental results from cortical neuronal cultures; again the results appear to confirm that neuronal cultures operate in close proximity of an instability line.

Taking advantage of the fact that the sizes of both \NS{} and quasi-orbits are found to be significantly correlated with the dynamic variable associated with \STD{} (available synaptic resources) just before the onset of the event, we developed a simple optimization method to infer, from the recorded activity, the characteristic time-scales of the putative fatigue mechanism at work. We first tested the method on the model, and then applied it to \textit{in-vitro} recordings; we could identify in several cases one or two long time-scales, ranging from few hundreds milliseconds to few seconds.

Weak or no correlations were found instead between avalanche sizes and the \STD{} dynamics; this suggests that avalanches originate from synaptic interaction which amplifies a wide spectrum of small fluctuations, and are mostly ineffective in eliciting a strong self-inhibition.

\section*{Models and Analysis}
\subsection*{Experimental data}
As originally described in \cite{eytan2006dynamics}, cortical neurons were obtained from newborn rats within 24 hours after birth, following standard procedures. Briefly, the neurons were plated directly onto a substrate-integrated multielectrode array (MEA). The cells were bathed in MEM supplemented with heat-inactivated horse serum (5\%), glutamine (0.5 mM), glucose (20 mM), and gentamycin (10 $\mu$g/ml) and were maintained in an atmosphere of $37^{\circ}$ C, 5\% CO2/95\% air in a tissue culture incubator as well as during the recording phases. The data analyzed here was collected during the third week after plating, thus allowing functional and structural maturation of the neurons. MEAs of 60 Ti/Au/TiN electrodes, 30 $\mu{}$m in diameter, and spaced 200 $\mu$m from each other (Multi Channel Systems, Reutlingen, Germany) were used. The insulation layer (silicon nitride) was pretreated with poly-D-lysine. All experiments were conducted under a slow perfusion system with perfusion rates of $\sim$100 $\mu$l/h. A commercial 60-channel amplifier (B-MEA-1060; Multi Channel Systems) with frequency limits of 1-5000 Hz and a gain of 1024$\times$ was used. The B-MEA-1060 was connected to MCPPlus variable gain filter amplifiers (Alpha Omega, Nazareth, Israel) for additional amplification. Data was digitized using two parallel 5200a/526 analog-to-digital boards (Microstar Laboratories, Bellevue, WA). Each channel was sampled at a frequency of 24000 Hz and prepared for analysis using the AlphaMap interface (Alpha Omega). Thresholds (8$\times$ root mean square units; typically in the range of 10-20 $\mu V$) were defined separately for each of the recording channels before the beginning of the experiment. The electrophysiological data is freely accessible for download at \url{marom.net.technion.ac.il/neural-activity-data/}.

\subsection*{Network rate dynamics}
A set of Wilson-Cowan-like equations \cite{wilson1972excitatory} for the spike-rate of the excitatory ($\nu_{E}$) and the inhibitory ($\nu_{I}$) neuronal populations lies at the core of our dynamic mean-field model:
\begin{eqnarray}
\label{eq.wilsoncowan}
\tau_E \, \dot{\nu}_E &=& -\big(\nu_E - \Phi(I_E)\big) \\
\tau_I \, \dot{\nu}_I &=& -\big(\nu_I - \Phi(I_I)\big), \nonumber
\end{eqnarray}
where $\tau_{E}$ and $\tau_{I}$ represent two characteristic times (of the order of few to few tens of $ms$), and $\Phi$ is the gain function of the input currents, $I_{E}$ and $I_{I}$, that in turn depend on $\nu_{E}$, $\nu_{I}$, and the synaptic efficacies. We chose $\Phi$ to be the transfer function of the leaky integrate-and-fire neuron under the assumptions of Gaussian, uncorrelated input of mean $\mu{}$ and infinitesimal variance $\sigma^2$ \cite{ricciardi1977diffusion}:
\begin{equation}
\label{eq.phi}
\Phi\big[\mu,\,\sigma^2\big] \equiv \Big[\sqrt{\pi}\,\tau_V\,\int_{\frac{V^{reset} - V^{rest} - \mu\,\tau_V}{\sqrt{\sigma^2 \, \tau_V}}}^{\frac{V^{thresh} - V^{rest} - \mu\,\tau_V}{\sqrt{\sigma^2 \, \tau_V}}}
\exp\big(s^2) \, \big[\mathrm{erf}\big(s) + 1\big] \, \mathrm{d}s + \tau_{refract} \Big]^{-1},
\end{equation}
where $\tau_V$ is the membrane time constant, $\tau_{refract}$ is a refractory period, and $V^{rest}$, $V^{reset}$, and $V^{thresh}$ are respectively the rest, the post-firing reset, and the firing-threshold membrane potential of the neuron (we assume the membrane resistance $R = 1$).

The model incorporates the non-instantaneous nature of synaptic transmission in its simplest form, by letting the $\nu{}$s being low-pass filtered by a single synaptic time-scale $\tilde{\tau}$:
\begin{equation}
\label{eq.nuTildeDyn}
\tilde{\tau} \, \dot{\tilde{\nu}} = \big(\nu - \tilde{\nu} \big).
\end{equation}
One can regard the variables $\tilde{\nu}$s as the instantaneous firing rates as seen by post-synaptic neurons, after synaptic filtering. The form of Eq.~\ref{eq.nuTildeDyn} and our choice of $\tilde{\tau}$ values (see Table~\ref{table.parameters}) implicitly neglects slow NMDA contributions and is restricted to AMPA and GABA synaptic currents. Thus, the input currents $I_{E}$ and $I_{I}$ in Eq.~\ref{eq.wilsoncowan} will be functions of the rates $\nu{}$s through these filtered rates; with reference to Eq.~\ref{eq.phi}, the model assumes the following form for the mean and the variance of the current $I_{E}$ (the expressions for $I_{I}$ are similarly defined):
\begin{eqnarray}
\mu_E & \equiv & c \, n_E \,  \tilde{\nu}_E \, \mathrm{w}_{\mathrm{exc}} \, J_{EE} \, r_E + \nonumber \\
&& c \, n_I \,  \tilde{\nu}_I \, \mathrm{w}_{\mathrm{inh}} \, J_{EI} + \nu_{ext} \, J_{ext} \label{eq.muAndSigma} \\
\sigma_E^2 & \equiv & c \, n_E \,  \tilde{\nu}_E \, \mathrm{w}^2_{\mathrm{exc}} \, \big(J_{EE}^2 + \sigma^2_{J_{EE}}\big) \, r_E^2  + \nonumber \\
&& c \, n_I \,  \tilde{\nu}_I \, \mathrm{w}^2_{\mathrm{inh}} \, \big(J_{EI}^2 + \sigma^2_{J_{EI}}\big) + \nu_{ext} \, \big(J_{ext}^2 + \sigma^2_{J_{ext}}\big), \nonumber
\end{eqnarray}
where the $n_E$ and $n_{I}$ are the number of neurons in the excitatory and inhibitory population respectively; $c$ is the probability of two neurons being synaptically connected; $J_{EE}$ ($J_{EI}$) is the average synaptic efficacy from an excitatory (inhibitory) pre-synaptic neuron to an excitatory one, $\sigma^2_{J}$ is the variance of the $J$-distribution; $\mathrm{w}_{\mathrm{exc}}$ and $\mathrm{w}_{\mathrm{inh}}$ are dimensionless parameters that we will use in the following to independently rescale excitatory and inhibitory synapses respectively. Finally, an external current is assumed in the form of a Poisson train of spikes of rate $\nu_{ext}$ driving the neurons in the network with average synaptic efficacy $J_{ext}$. In Eq.~\ref{eq.muAndSigma} $r_{E}(t)$ ($0 < r_{E} < 1$) is the fraction of synaptic resources available at time $t$ for the response of an excitatory synapse to a pre-synaptic spike; the evolution of $r_{E}$ evolves according to the following dynamics, which implements the effects of short-term depression (\STD{}) \cite{tsodyks1997neural,tsodyks1998neural} into the network dynamics:
\begin{equation}
\label{eq.rStd}
\tau_{\mathrm{STD}} \, \dot{r}_{E} = (1 - r_{E}) - u_{\mathrm{STD}}\,r_{E}\,\tau_{\mathrm{STD}}\, \tilde{\nu}_E ,
\end{equation}
where $0 < u_{\mathrm{STD}} < 1$ represents the (constant) fraction of the available synaptic resources consumed by an excitatory postsynaptic potential, and $\tau_{\mathrm{STD}}$ is the recovery time for the synaptic resources.

Finally, for a network of $n$ neurons, we introduce finite-size noise by assuming that the signal the synapses integrate in Eq.~\ref{eq.nuTildeDyn} is a random process $\nu_{n}$ of mean $\nu{}$; in a time-bin $\mathrm{dt}$, we expect the number of action potentials fired to be a Poisson variable of mean $n\,\nu(t) \, \mathrm{dt}$; Eq.~\ref{eq.nuTildeDyn} will thus become:
\begin{eqnarray}
\label{eq.nuTildeDynNoisy}
\tilde{\tau} \, \dot{\tilde{\nu}} &=& \big(\nu_n - \tilde{\nu} \big) \nonumber \\
\nu_n &\equiv& \frac{\mathrm{Poisson}[n\,\nu\,\mathrm{dt}]}{n\,\mathrm{dt}} .
\end{eqnarray}

Putting all together, the noisy dynamic mean-field model is described by the following set of (stochastic) differential equations:
\begin{equation}
\label{eq.noisyDynamicMF}
\left\{
\begin{array}{ccl}
\tau_E \, \dot{\nu}_E &=& \Phi\big(\mu_E,\,\sigma^2_E\big) - \nu_E \\
\tau_I \, \dot{\nu}_I &=& \Phi\big(\mu_I,\,\sigma^2_I\big) - \nu_I \\
\tilde{\tau}_E \, \dot{\tilde{\nu}}_E &=& \nu_{n_E} - \tilde{\nu}_E \\
\tilde{\tau}_I \, \dot{\tilde{\nu}}_I &=& \nu_{n_I} - \tilde{\nu}_I \\
\tau_{\mathrm{STD}} \, \dot{r}_E &=& (1 - r_E) - u_{\mathrm{STD}} \, \tau_{\mathrm{STD}} \, r_E \, \tilde{\nu}_E
\end{array}
\right.
\end{equation}
complemented by Eqs.~\ref{eq.phi}, \ref{eq.muAndSigma}, and \ref{eq.nuTildeDynNoisy}. The values of all the fixed network parameters are shown in Table~\ref{table.parameters}. Since we will compare the dynamics of networks of different sizes, we scale the connectivity with network size in order to keep invariant the mean field equations: we hold the number of synaptic connection per neuron constant by rescaling, with reference to Eq.~\ref{eq.muAndSigma}, the probability of connection $c$ so that $c \, n_E$ and $c \, n_I$ are kept constant to the reference values that can be deduced from Table~\ref{table.parameters}.

\begin{table}[!ht]
\caption{\bf{Network parameters.}} 
\centering 
\begin{tabular}{|c|c|}
\hline
Parameter & Value \\ [0.5ex] 
\hline 
$n_E$ / $n_I{}$ & 160 / 40 \\ \hline
$\tau_V$ / $\tau_{refract}$ & $20$ / $2$ ms \\ \hline
$V^{rest}$ / $V^{firing}$ & $-70$ / $-55$ mV \\ \hline
$c$ & 0.25 \\ \hline
$J_{EE} \pm \sigma_{J_{EE}}$ & $0.809 \pm 0.202$ mV  \\ \hline
$J_{IE} \pm \sigma_{J_{IE}}$ & $1.23 \pm 0.307$ mV \\ \hline
$J_{EI} \pm \sigma_{J_{EI}}$ & $-0.340 \pm 0.0850$ mV \\ \hline
$J_{II} \pm \sigma_{J_{II}}$ & $-0.358 \pm 0.0894$ mV \\ \hline
$J_{ext} \pm \sigma_{J_{ext}}$ & $0.416 \pm 0.104$ mV \\ \hline
$\nu_{ext}$ & $1.25$ kHz \\ \hline
$\tilde{\tau}_E$ / $\tilde{\tau}_I$ & $10$ / $2$ ms \\ \hline
$\tau_r$ & $800$ ms \\ \hline
$u_{\mathrm{STD}}$ & 0.2 \\ \hline 
$\tau_E = \tau_I$ & $20$ ms \\
\hline 
\end{tabular}
\label{table.parameters} 
\end{table}

Spike-frequency adaptation (\SFA{}) (not present in simulations unless where explicitly stated) is introduced by subtracting a term to the instantaneous mean value of the $I_{E}$ current:
\begin{equation}
\label{eq.muEMinusC}
\mu_E \rightarrow \mu_E - g_{\mathrm{SFA}}\,c_E(t)
\end{equation}
proportional to the instantaneous value of the variable $c_E$, that simply integrates $\nu_{n_E}$:
\begin{equation}
\label{eq.cDynamics}
\tau_{\mathrm{SFA}} \, \frac{\mathrm{d} c_E}{\mathrm{d}t} = -c_E + \nu_{n_E} ,
\end{equation}
with a characteristic time $\tau_{\mathrm{SFA}}$. This additional term aims to model an after-hyperpolarization, $\mathrm{Ca}^{2+}$-dependent $\mathrm{K}^+$ current \cite{wang1998calcium,ermentrout1998linearization}. In this sense, $c_E$ can be interpreted as the cytoplasmic calcium concentration [$\mathrm{Ca}^{2+}$]), whose effects on the network dynamics are controlled by the value of the ``conductance'' $g_{\mathrm{SFA}}$.

Simulations are performed by integrating the stochastic dynamics with a fixed time step $\mathrm{dt} = 0.25$ ms.

In the following, by ``spike count'' we will mean the quantity $\nu(t) \, n \, \mathrm{dt}$.

\subsection*{Network events detection}
For the detection of network events (\NS{}s, quasi-orbits, and avalanches) we developed a unified approach based on Hidden Markov Models (HMM) \cite{baum1966statistical}. Despite HMM have been widely used for temporal pattern recognition in many different fields, to our knowledge few attempts have been made to use them in the context of interest here \cite{chen2009discrete,tokdar2010detection}. For the purpose of the present description, we just remind that a HMM is a stochastic system that evolves according to Markov transitions between ``hidden'', i.e. unobservable, states; at each step of the dynamics the visible output depends probabilistically on the current hidden state. Such models can be naturally adapted to the detection of network events, the observations being the number of detected spikes per time bin, and the underlying hidden states, between which the system spontaneously alternates, being associated with high or low network activity (`network event - no network event'). A standard optimization procedure adapts then the HMM to the recorded activity sample by determining the most probable sequence of hidden states given the observations.

The two-step method we propose is based on HMM, has no user-defined parameters, and automatically adapts to different conditions. 

In the first step, the algorithm finds the parameters of the two-state HMM (one low-activity state, representing the quasi-quiescent periods, and one high-activity state, associated with network events) that best accounts for a given sequence of spike counts -- the visible states in the HMM; such fitting is performed through the Baum-Welch algorithm \cite{baum1966statistical}. In the second step, the most probable sequence for the two alternating hidden levels, given the sequence of spike counts and the fitted parameters, is found through the Viterbi algorithm. Network events are identified as the periods of dominance of the high activity hidden state.

In order to retain only the most significant events a minimum event duration is imposed; such threshold is self-consistently determined as follows. The Viterbi algorithm is also applied to a ``surrogate'' time-series obtained by randomly shuffling the original one, thereby generating a set of ``surrogate'' events. The purpose is to determine the desired minimum event duration from the high duration tail of surrogate events (which, by construction, come from a time-series with no real temporal structure). Since the high duration distribution tail is found to be roughly exponential, we fit such tail by considering only the surrogate events of duration larger than the 75th percentile. Then, from the fitted exponential, we compute the duration value such that the probability of durations greater than this value is $P(\mathrm{surrogate}) = 10^{-3}$. In other words, we set the threshold on minimum duration of detected events to the duration of exceptionally long ($P < 10^{-3}$) surrogate events.

As already remarked, we used essentially the same algorithm for detecting \NS{}/quasi-orbits and avalanches. The only significant difference is that, in the case of avalanches, the emission probability of the low-activity hidden state is kept fixed during the Baum-Welch algorithm to $p(n) \simeq \delta_{n0}$ ($\delta_{ij}$ is the Kronecker delta; $p(n)$ is the probability of emitting $n$ spikes in a time-bin). Thus the lower state is constrained to a negligible probability of outputting non-zero spike-counts, conforming to the intuition that in between avalanches the network is (almost) completely silent. More precisely, we set $p(1) = 10^{-6} \, \langle n \rangle$, where $\langle n \rangle$ is the average number of spikes that the network emits during a time-bin $\mathrm{dt}$. After the modified Baum-Welch first step, avalanches are determined, as above, by applying the Viterbi algorithm; no threshold is applied in this case, neither to the avalanche duration nor to its size.

The proposed procedures introduce three arbitrary parameters: the time bin $\mathrm{dt}$, the probability $P(\mathrm{surrogate})$ for network spikes and quasi-orbits, and the probability $p(1)$. To test the robustness of the algorithms, we varied these parameters over ample ranges: $\mathrm{dt}$ between 0.25 and 8 ms; $P(\mathrm{surrogate})$ between $10^{-2}$ and $10^{-4}$; $p(1)$ between $10^{-8}$ and $10^{-4}$. We found that avalanche size distributions are virtually unaffected under variations of $p(1)$, and only mildly affected for the largest $\mathrm{dt}$ explored; higher values of $P(\mathrm{surrogate})$ lead, as expected, to detect a larger number of small quasi-orbits, yet these additional events do not alter the overall shape of the size distribution predicted by the theory (see next section); on the other hand, a large number of very small quasi-orbits does have a detrimental effect on the correlation results reported in Section ``Inferring the time-scales''.

Simulations and data analysis have been performed using custom-written mixed C++/
MATLAB (version R2013a, Mathworks, Natick, MA) functions and scripts.

\subsection*{Size distribution for quasi-orbits and network spikes}
The non-linear rate model described above can show a wide repertoire of dynamical patterns, as for example multiple stable fixed points and large, quasi-periodic oscillations. As we will show, for sufficiently excitable networks, a stable state of asynchronous activity (fixed point) is destabilized, in favor of stable global oscillations. Finite size noise probes differently network's excitability at different distances from such instability. Before global oscillations become stable (in the infinite network limit), the network's highly non-linear reaction to its own fluctuations can ignite large, relatively stereotyped, ``network spikes''. Also, in the proximity of the oscillatory (Hopf) instability, noise can promote ``quasi orbits'', \textit{i.e.}, transient departures from the fixed point which develop on time-scales dictated by the upcoming oscillatory instability, of which they are precursors. Under a linear approximation, the probability distribution of the amplitude $l$ of these quasi-orbits can be explicitly derived as explained in the following.

Consider a generic planar linear dynamics with noise:
\begin{equation}\label{eq.LinearSystem0}
\mathbf{\dot{z}} = \mathcal{A}\,\mathbf{z} + \sigma \, \boldsymbol{\xi} ,
\end{equation}
where $\mathcal{A}$ is $2\times 2$ real matrix, and
$\boldsymbol{\xi} = (\xi(t),\,0)$ is a white noise with $\langle
\xi(t)\xi(t') \rangle = \delta(t-t')$. We here
assume that the system is close to a Hopf
bifurcation; in other words that the matrix $\mathcal{A}$ has
complex-conjugated eigenvalues $\lambda_\pm = \Re \lambda +
\mathrm{i}\,\Im \lambda$, with $\Re \lambda < 0$ and $|\Re \lambda| \ll \Im \lambda$.

By means of a linear transformation, the system can be rewritten as:
\begin{eqnarray}
\dot{x} &=& \Re \lambda \, x - \Im \lambda \, y + \sigma_x \, \xi \nonumber \\
\dot{y} &=& \Im \lambda \, x + \Re \lambda \, y + \sigma_y \, \xi , \nonumber \\
\end{eqnarray}
with $\sigma_x$ and $\sigma_y$ constants determined by the coordinate transformation. 
Making use of It\={o}'s lemma to write:
\begin{eqnarray}
\dot{x^2} &=& 2 \, \Re \lambda \, x^2 - 2\, \Im \lambda \, x \, y + \sigma_x^2 + 2 \, x \, \sigma_x \, \xi \nonumber \\
\dot{y^2} &=& 2 \, \Im \lambda \, x \, y + 2 \, \Re \lambda \, y^2 + \sigma_y^2 + 2 \, y \, \sigma_y \, \xi \nonumber ,
\end{eqnarray}
and summing the previous two equations, we find for the square radius $l^2 \equiv x^2 + y^2$ the dynamics:
\begin{equation}\label{eq.dr2}
\dot{l^2} = 2 \, \Re \lambda \, l^2 + \sigma'^2 +
2 \, \big(\sigma_x \, x + \sigma_y \, y\big)\,\xi ,
\end{equation}
with $\sigma'^2 \doteq \sigma_x^2+\sigma_y^2$. 

As long as $\Im \lambda \gg |\Re \lambda|$, it is physically sound to make the approximation:
\begin{equation}\label{eq.xrtheta}
(x(t), \, y(t)) = l(0) \, \big(\cos(\Im \lambda\,t + \phi),\,\sin(\Im \lambda\,t + \phi)\big) ,
\end{equation}
for $0 \leq t \leq T = 2\,\pi / \Im \lambda$ and then to average the variance of the noise over such period to get:
\begin{eqnarray*}
&&\frac{l(0)^2}{T}\,\int_0^T\,\Big[\sigma_x\,\cos(\Im \lambda\,t + \phi)+\sigma_y\,\sin(\Im \lambda\,t + \phi)\Big]^2\mathrm{d}t = \\
&=& l(0)^2 \, \frac{\sigma_x^2+\sigma_y^2}{2} = \frac{l(0)^2 \, \sigma'^2}{2}.
\end{eqnarray*}
in order to rewrite Eq.~(\ref{eq.dr2}) as:
\begin{equation}\label{eq.dr2Approx}
\dot{l^2} = 2 \, \Re \lambda \, l^2 + \sigma'^2 + \sqrt{2} \,l \, \sigma' \, \xi .
\end{equation}

Such stochastic differential equation is associated with the Fokker-Planck equation:
\begin{eqnarray}\label{eq.FPr2}
&&\partial_t\,p(l^2,t) = -\partial_{l^2}\,\big[2\,\Re \lambda\,l^2 +
\sigma'^2]\,p(l^2,t) + \\
&+& \sigma'^2\,\partial^2_{l^2}\,l^2\,p(l^2,t) \equiv L_{l^2}\,p(l^2,t) \nonumber
\end{eqnarray}
that admits an exponential distribution as stationary solution:
\begin{equation}\label{eq.Pssr2}
p_{ss}(l^2) = \frac{2\,|\Re \lambda|}{\sigma'^2}\,\exp\big(
-\frac{2\,|\Re \lambda|\,l^2}{\sigma'^2}\big) ,
\end{equation}
that is, a Rayleigh distribution for $l$:
\begin{equation}\label{eq.Pssr1}
p_{ss}(l) = \frac{4\,|\Re \lambda|}{\sigma'^2}\,l\,\exp\big(
-\frac{2\,|\Re \lambda|\,l^2}{\sigma'^2}\big) .
\end{equation}

On the other hand, we found a correlation between $l$ (the maximal departure from the low-activity fixed point) and the duration of the quasi-orbit. Therefore the size of the quasi-orbit (the `area' below the firing rate time profile during the excursion from the fixed point) is expected to scale as $l^2$, so that it should be exponentially distributed.

For network spikes we do not have a theoretical argument to predict the shape of the size distribution, however empirically a (left-truncated) Gaussian distribution proved to be roughly adequate. Since we expect that quasi-orbits and \NS{} contribute with different weights for varying excitatory/inhibitory balance, we adopted the following form for the overall distribution of network event size to fit experimental data:
\begin{equation}
\label{eq.pSizeNSQO}
p(x) =  \frac{p_0}{\tau_0} \, \exp\big(-\frac{(x - x_0)}{\tau_0} \big) +
\frac{1 - p_0}{\sqrt{2\,\pi} \, \sigma_1} \, \exp\big(-\frac{(x - m_1)^2}{2 \, \sigma_1^2} \big) .
\end{equation}
The parameters of the two distributions and their relative weight $0 \leq p_0 \leq 1$ are estimated by minimizing the log-likelihood on the data.
A threshold for the event size is determined as the value having equal probability of being generated by either the exponential or the normal distribution. In the following, \NS{}s are defined as events having size larger than this threshold. In those cases in which a threshold smaller than the peak of the normal distribution could not be determined, no threshold was set.

\section*{Results}
In the following, we will study a stochastic firing-rate model and make extensive comparison of its dynamical behavior with the activity of \textit{ex-vivo} networks of cortical neurons recorded through a 60-channel multielectrode array.

The first question we want to answer is how the excitation-inhibition balance affects network dynamics. Starting from the statistics of network spikes (\NS) we show that it is well described by a single variable measuring the distance from an oscillatory instability of the dynamics. We then study in the model the effects of finite-size fluctuations on the statistics of \NS.

Then, taking advantage of new detection algorithm we introduce (see Models and Analysis), we recognize the presence of a spectrum of network events, including three families: \NS{}, ``quasi-orbits'', and avalanches. The predicted size distribution of quasi-orbits , exponential component in Eq.~\ref{eq.pSizeNSQO}, is confirmed by simulations and recovered in experimental data analysis. We investigate how the different network events characterize in various proportions the network dynamics depending on the excitatory-inhibitory balance; experimental data offer an interesting match with model findings, compatible with \textit{ex-vivo} network being typically slightly below the oscillatory instability.

Finally we introduce a simple procedure to infer the time-scales of putative slow self-inhibitory mechanisms underlying the occurrence of network events. The inference is obtained based on knowledge of the firing activity alone; this makes the method interesting for analysis of experimental data, as we show through exemplary results.

The stochastic firing-rate model consists of two populations of neurons, one excitatory and one inhibitory, interacting through effective synaptic couplings; excitatory synaptic couplings follow a dynamics mimicking short-term depression (described after the Tsodyks-Markram model, \cite{tsodyks1997neural}). We adopted the transfer function of the leaky integrate-and-fire neuron subject to white-noise current with drift \cite{ricciardi1977diffusion} as the single population input-output function; moreover the activity of each population is made stochastic by adding realistic finite-size noise. Working with a noisy mean field model allows in principle to easily sweep through widely different network sizes and, more importantly, allows us to perform the stability analysis.

\begin{figure}[h!]
\begin{center}
\setlength{\unitlength}{\textwidth}
\begin{picture}(1,0.55)

  \put(0.024,0)
  {
    \epsfig
    {
    file=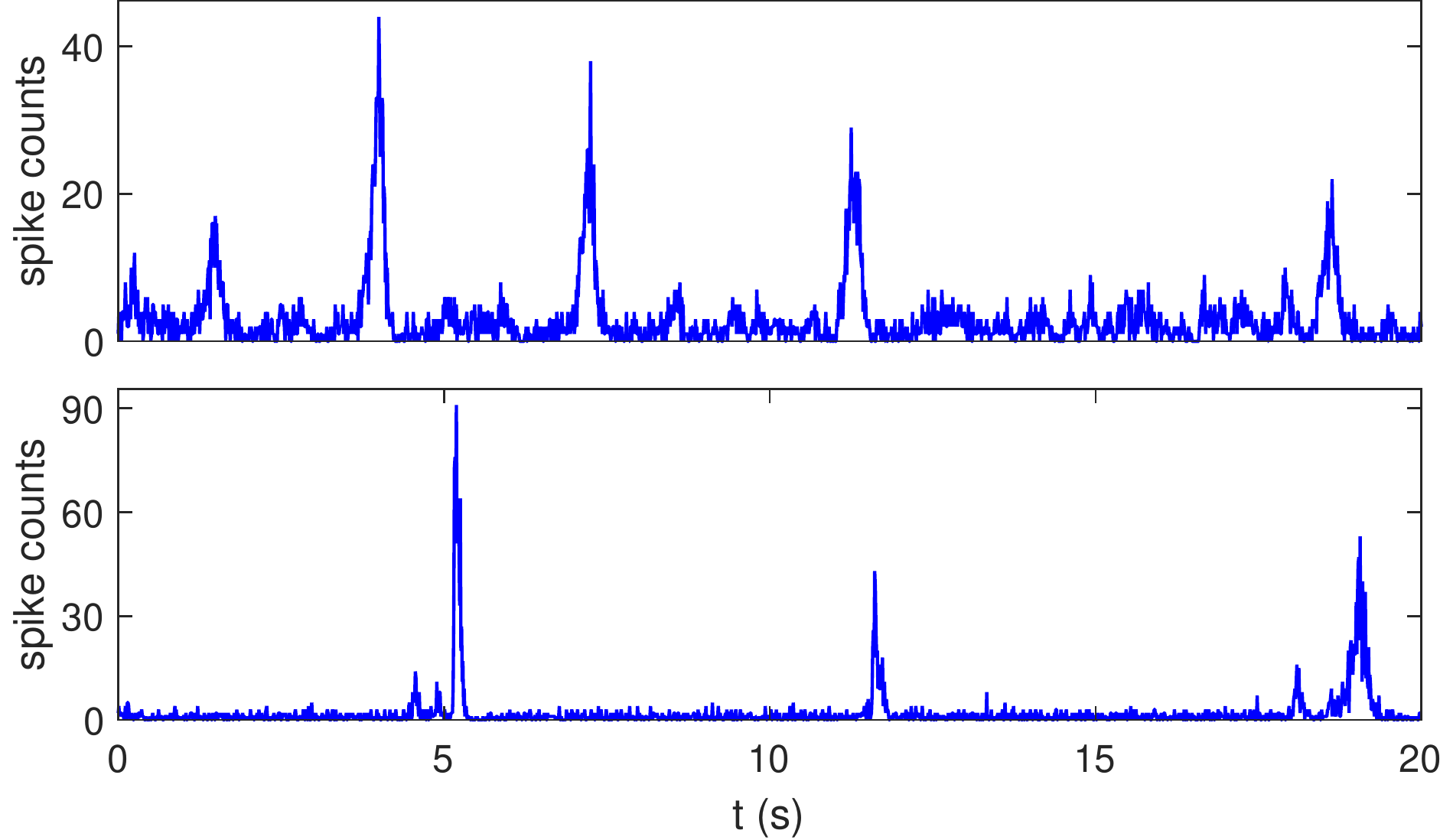,
    width=0.95\unitlength,
    }
  }
  \put(0.16,0.47){\makebox{\Large\textbf{A}}}
  \put(0.16,0.235){\makebox{\Large\textbf{B}}}
\end{picture}
\caption{{\bf Time course of the network firing rate.}
Panel A: noisy mean-field simulations; panel B: \textit{ex-vivo} data. Random large excursions of the firing rate (network spikes and quasi-orbits) are clearly visible in both cases.}
\label{figure1}
\end{center}
\end{figure}

To start the exploration that follows, we chose a reference working point where the model's dynamics has a low-rate fixed point ($2-4$ Hz) just on the brink of an oscillatory instability or, in other words, where the dominant eigenvalue $\lambda{}$ of the dynamics, linearized around the fixed point, is complex with null real part. The model network (Fig.~\ref{figure1}, panel A) shows in proximity of this point a dynamical behavior qualitatively similar, in terms of population spikes, to what is observed in \textit{ex-vivo} neuronal networks (Fig.~\ref{figure1}, panel B). 

\subsection*{Excitation-inhibition balance and network spike statistics}
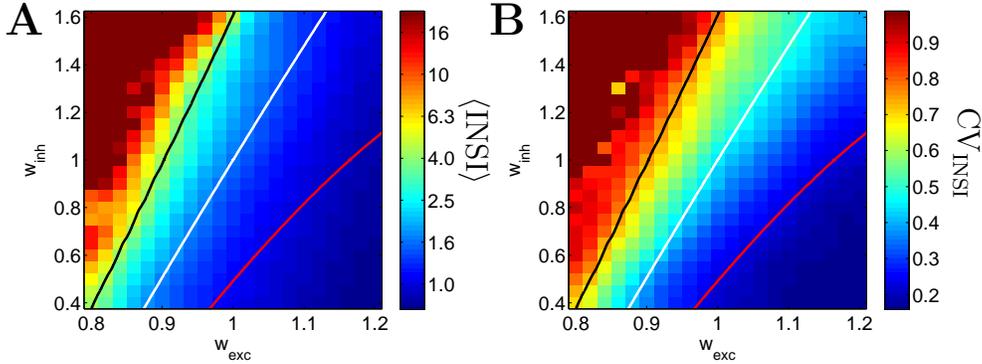
\begin{figure}[h!]
\begin{center}
\setlength{\unitlength}{\textwidth}
\begin{picture}(1,0.39)
  \put(0.02,0)
  {
    \epsfig
    {
    file=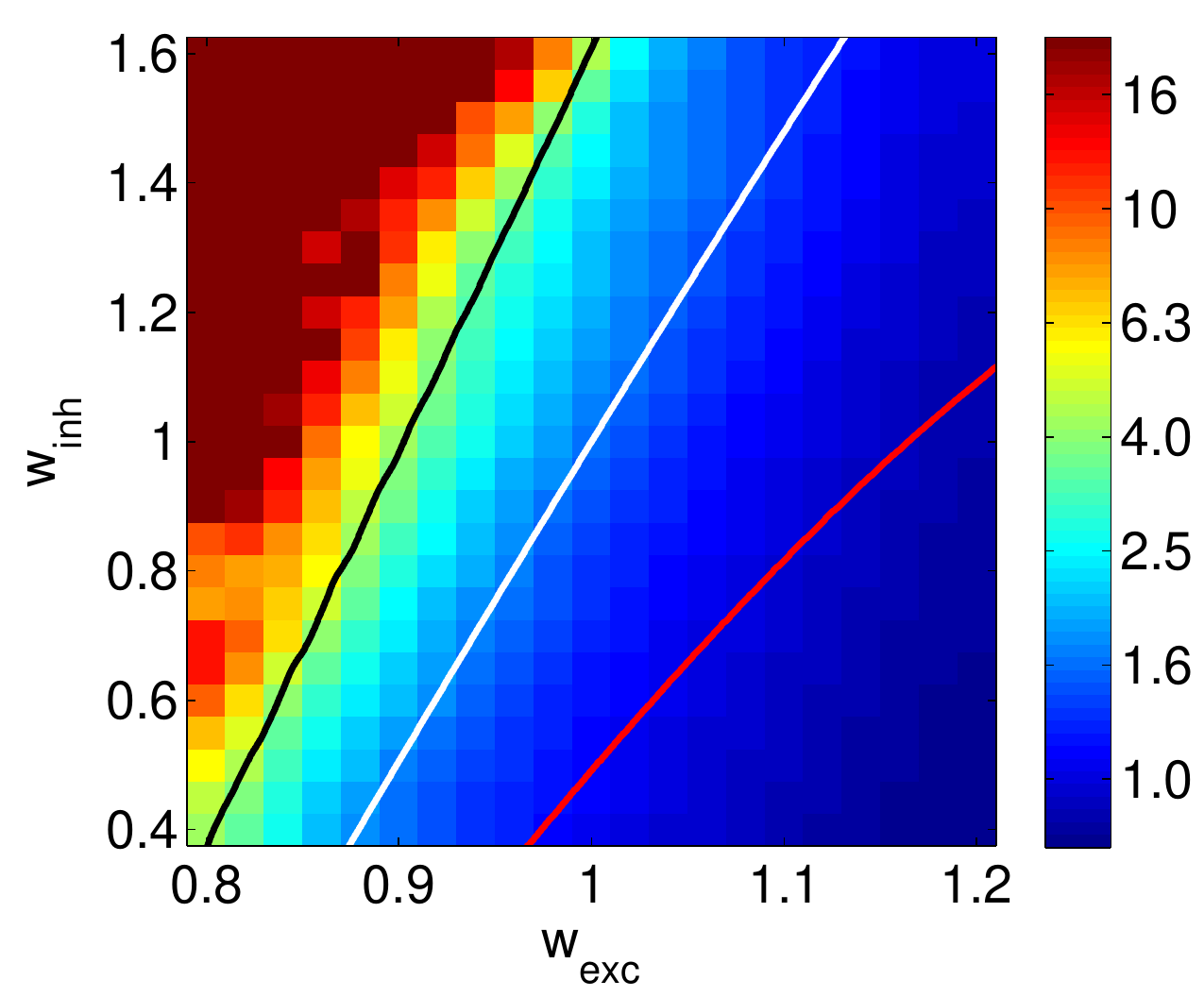,
    width=0.43\unitlength,
    }
  }
  \put(0.49,0)
  {
    \epsfig
    {
    file=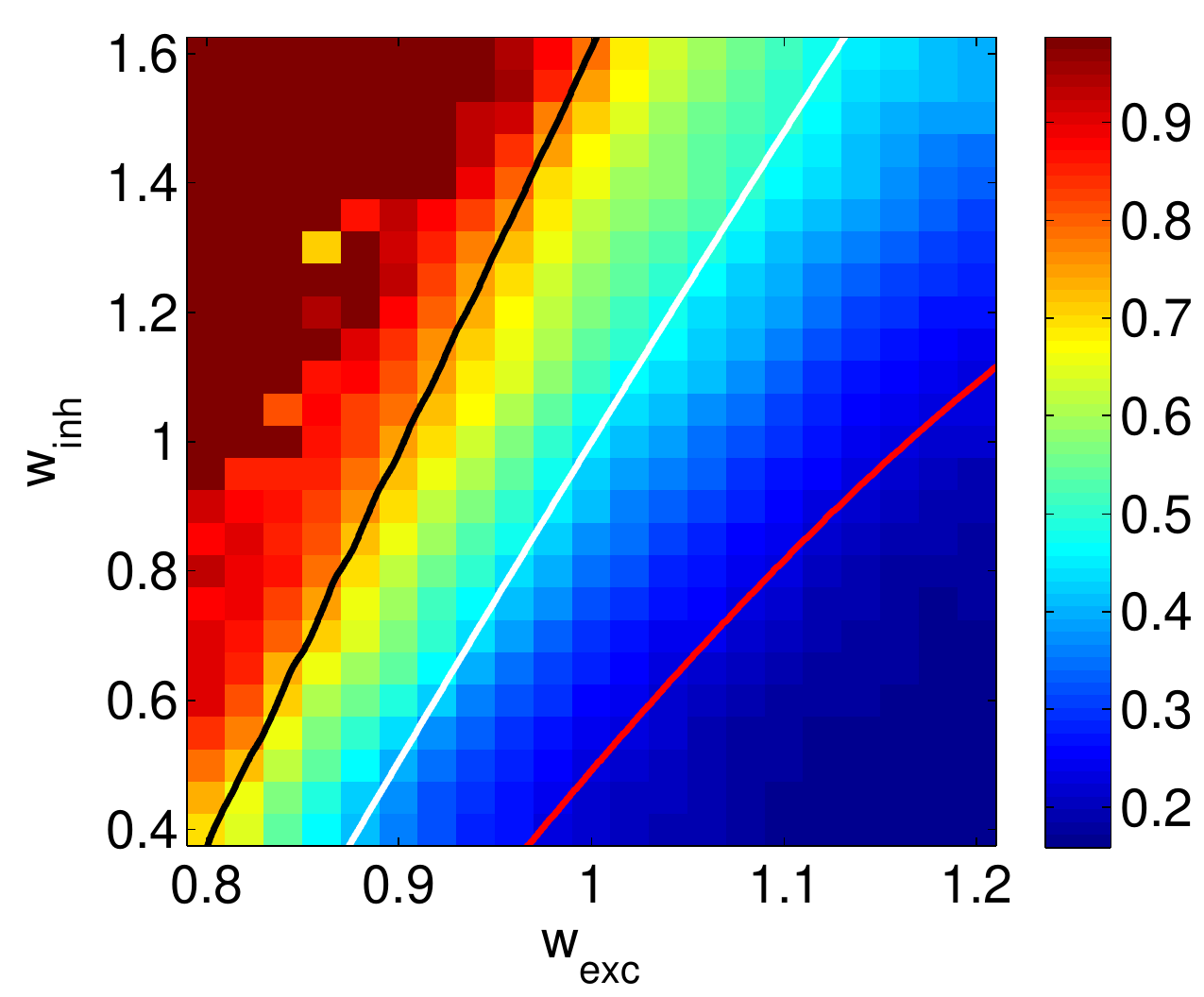,
    width=0.43\unitlength,
    }
  }
  \put(0.02,0.32){\makebox{\Large\textbf{A}}}
  \put(0.49,0.32){\makebox{\Large\textbf{B}}}
  
  \put(0.46,0.26){\rotatebox{270}{$\langle \mathrm{INSI} \rangle$}}
  \put(0.94,0.25){\rotatebox{270}{$\mathrm{CV}_{\mathrm{INSI}}$}}
\end{picture}
\caption{{\bf Inter-network-spike interval (INSI) statistics in the noisy mean-field model, for varying levels of excitation ($\mathrm{w}_{\mathrm{exc}}$) and inhibition ($\mathrm{w}_{\mathrm{inh}}$).}
Panel A: $\langle \mathrm{INSI} \rangle$ (the scale is in seconds); panel B: coefficient of variation of INSI ($\mathrm{CV}_{\mathrm{INSI}}$). For high net excitation (bottom-right quadrant) short-term depression plays a determinant role in generating frequent and regular (low $\mathrm{CV}_{\mathrm{INSI}}$) \NS{}s; for weak excitability (upper-left quadrant) random fluctuations are essential for the generation of rare, quasi-Poissonian \NS{}s ($\mathrm{CV}_{\mathrm{INSI}} \simeq 1$). The solid lines are isolines of the real part $\Re \lambda{}$ of the dominant eigenvalue of the mean-field dynamics' Jacobian; white line: $\Re \lambda{} = 0$ Hz; red line: $\Re \lambda{} = 3.5$ Hz; black line: $\Re \lambda{} = -3.5$ Hz. Note how such lines roughly follow isolines of $\langle \mathrm{INSI} \rangle$ and $\mathrm{CV}_{\mathrm{INSI}}$.}
\label{figure2}
\end{center}
\end{figure}

As the relative balance of excitation and inhibition is expected to be a major determinant of \NS{} statistics we investigated first, for spontaneous \NS{}s, how the inter-\NS{} intervals (INSI) and their regularity (as measured by the coefficient of variation, $\mathrm{CV}_{\mathrm{INSI}}$) depend on such balance. In Fig.~\ref{figure2} we report the average INSI (left panel) and $\mathrm{CV}_{\mathrm{INSI}}$ (right panel) in the plane ($\mathrm{w}_{\mathrm{exc}}$, $\mathrm{w}_{\mathrm{inh}}$) of the excitatory and inhibitory synaptic efficacies ($J_{EE} \rightarrow w_E \, J_{EE}$, $J_{IE} \rightarrow w_E \, J_{IE}$, $J_{EI} \rightarrow w_I \, J_{EI}$, $J_{II} \rightarrow w_I \, J_{II}$, see Eq.~\ref{eq.muAndSigma}). Starting from the center of this plane ($\mathrm{w}_{\mathrm{exc}} = 1$, $\mathrm{w}_{\mathrm{inh}} = 1$) and moving along the horizontal axis, all the excitatory synapses of the network are multiplied by a factor $\mathrm{w}_{\mathrm{exc}}$: moving right, the total excitation of the network increases ($\mathrm{w}_{\mathrm{exc}} > 1$), toward left it decreases ($\mathrm{w}_{\mathrm{exc}} < 1$). Along the vertical line, instead, all the inhibitory synapses are damped (moving downward, $\mathrm{w}_{\mathrm{inh}} < 1$) or strengthened (going upward, $\mathrm{w}_{\mathrm{inh}} > 1$).

It is clearly seen that both $\langle \mathrm{INSI} \rangle{}$ and $\mathrm{CV}_{\mathrm{INSI}}$ are approximately distributed in the plane along almost straight lines of equal values: for a chosen $\langle \mathrm{INSI} \rangle$ or $\mathrm{CV}_{\mathrm{INSI}}$ one can trade more excitation for less inhibition keeping the value constant, suggesting that, at this level of approximation, a measure of net synaptic excitation governs the \NS{} statistics.
Besides, not surprisingly, for high net excitation \NS{}s are more frequent ($ \sim 1$ Hz) and quasi-periodic (low $\mathrm{CV}_{\mathrm{INSI}}$), due to the fact that the \STD{} recovery time determines quasi-deterministically when the network is again in the condition of generating a new \NS{}. Weak excitability, on the other hand, leads to rare \NS{}s, approaching a Poisson statistics ($\mathrm{CV}_{\mathrm{INSI}} \simeq 1$), since excitability is so low that fluctuations are essential for recruiting enough activation to elicit a \NS{}, with \STD{} playing little or no role at the ignition time; below an ``excitation threshold'', \NS{}s disappear.

The solid lines in Fig.~\ref{figure2} are derived from the linearization of the 5-dimensional dynamical system (see Eq.~\ref{eq.noisyDynamicMF}), and are curves of iso-$\Re \lambda{}$, where $\lambda$ is the dominant eigenvalue of the Jacobian: $\Re \lambda{} = 0$ Hz (white line, signaling a Hopf bifurcation in the corresponding deterministic system), $\Re \lambda{} = 3.5$ Hz (red line), and $\Re \lambda{} = -3.5$ Hz (black line). Values of $\mathrm{CV}$ found in typical cultured networks are close to model results near the bifurcation line $\Re \lambda{} = 0$ Hz. We observe, furthermore, that such lines roughly follow iso-$\langle \mathrm{INSI} \rangle$ and iso-$\mathrm{CV}_{\mathrm{INSI}}$ curves, suggesting that a quasi one-dimensional representation might be extracted. 

\begin{figure}[h!]
\begin{center}
\setlength{\unitlength}{\textwidth}
\begin{picture}(1,0.35)

  \put(-0.01,0)
  {
    \epsfig
    {
    file=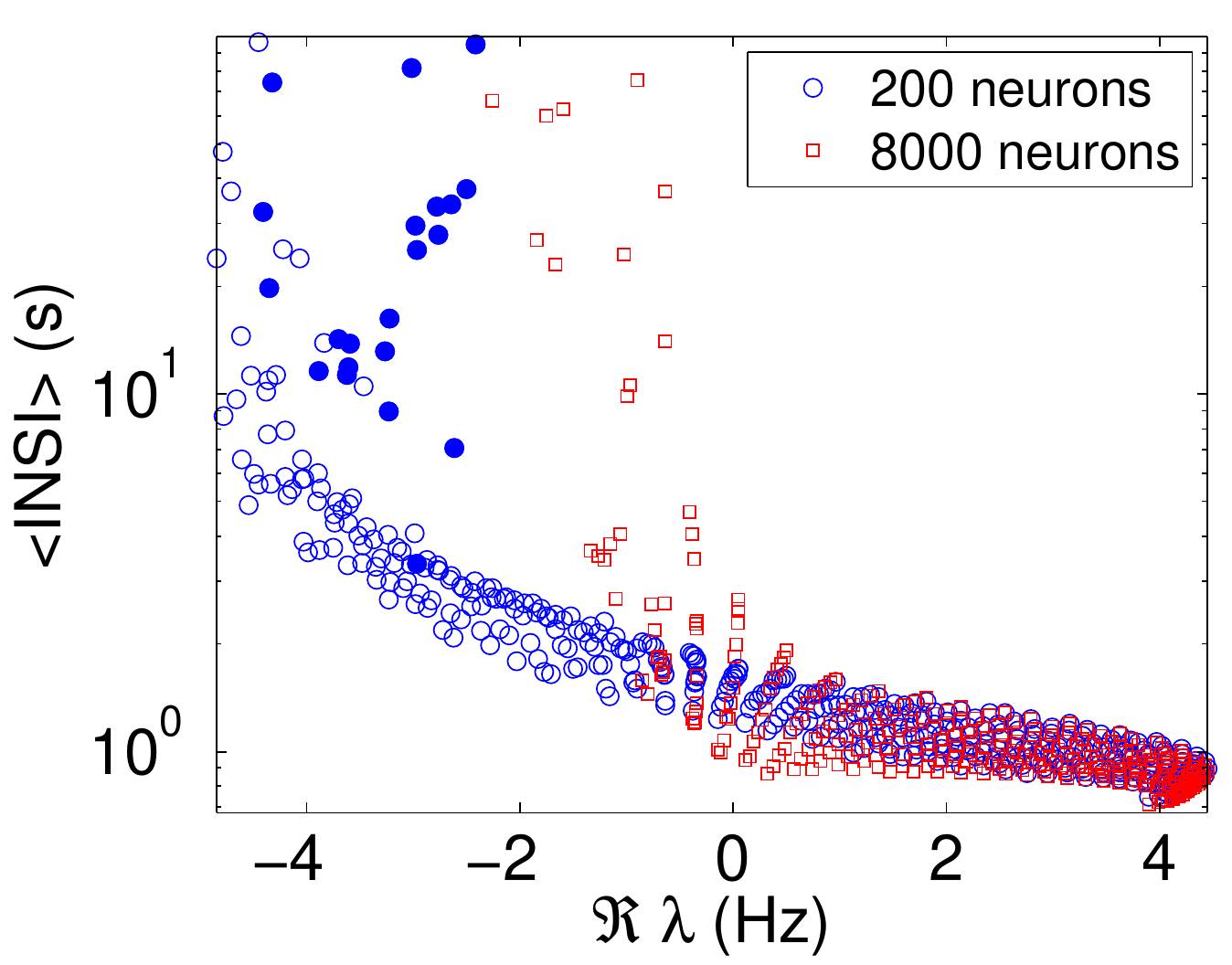,
    width=0.45\unitlength,
    }
  }
  \put(0.5,0)
  {
    \epsfig
    {
    file=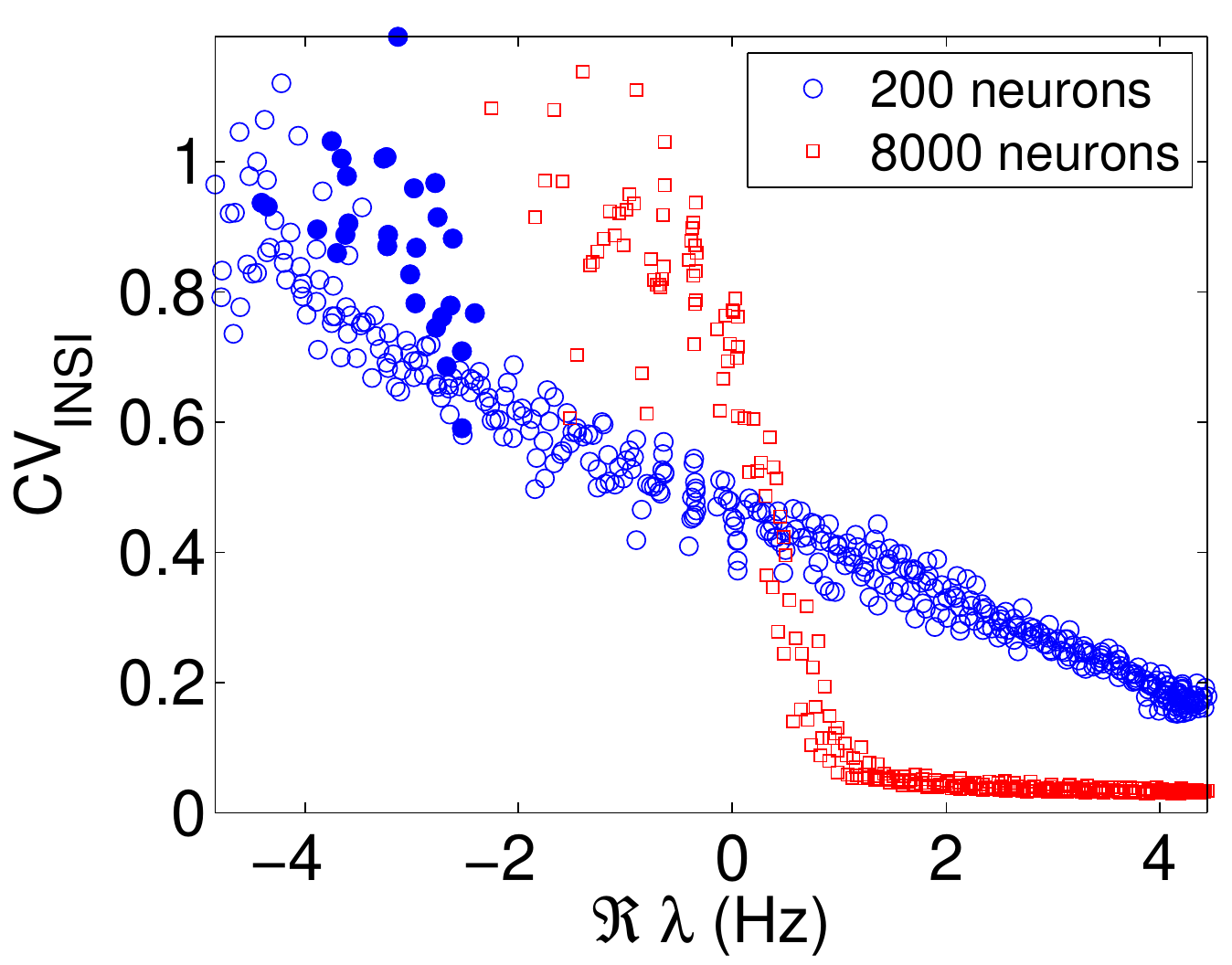,
    width=0.45\unitlength,
    }
  }
  \put(-0.01,0.315){\makebox{\Large\textbf{A}}}
  \put(0.51,0.315){\makebox{\Large\textbf{B}}}
\end{picture}
\caption{{\bf Stability analysis of the linearized dynamics captures most of the variability in the inter-network-spike interval (INSI) statistics.}
$\langle \mathrm{INSI} \rangle{}$ (panel A) and $\mathrm{CV}_{\mathrm{INSI}}$ (panel B) \textit{vs} the real part $\Re \lambda$ of the dominant eigenvalue of the Jacobian of the linearized dynamics, for two networks that are pointwise identical in the excitation-inhibition plane, except for their size (circles: 200 neurons, as in Fig.~\ref{figure2}; squares: 8000 neurons). The data points almost collapse on 1-D curves when plotted as functions of $\Re \lambda$, leading effectively to a ``quasi one-dimensional'' representation of the INSI statistics in the $(\mathrm{w}_{\mathrm{exc}}, \, \mathrm{w}_{\mathrm{inh}})$-plane.
The region in which the INSIs are neither regular ($\mathrm{CV}_{\mathrm{INSI}} \sim 0$) nor completely random ($\mathrm{CV}_{\mathrm{INSI}} \simeq 1$), as typically observed in experimental data, shrinks for larger networks. The filled circles mark a null imaginary part $\Im \lambda$.}
\label{figure3}
\end{center}
\end{figure}

We show in Fig.~\ref{figure3} $\langle \mathrm{INSI} \rangle$ (panel A) and $\mathrm{CV}_{\mathrm{INSI}}$ (panel B) against $\Re\lambda$ for the same networks (circles) of Fig.~\ref{figure2}, and for a set of larger networks ($N = 8000$ neurons, squares) that are otherwise identical to the first ones, pointwise in the excitation-inhibition plane (the average number of synaptic connections per neuron for the larger networks is kept constant to the value used in the original, smaller ones, as explained in Models and Analysis) The difference in size amounts, for the new, larger networks, to weaker endogenous noise entering the stochastic dynamics of the populations' firing rates (see Eq.~\ref{eq.nuTildeDynNoisy}, second line). The points are seen to approximately collapse onto lines for both sets of networks, thus confirming $\Re \lambda{}$ as the relevant control quantity for $\langle \mathrm{INSI} \rangle$ and $\mathrm{CV}_{\mathrm{INSI}}$. It is seen that, for the smaller networks, $\langle \mathrm{INSI} \rangle $ and $\mathrm{CV}_{\mathrm{INSI}}$ depend smoothly on $\Re\lambda$, due to finite-size effects smearing the bifurcation. Also note the branch of points (filled circles) for which $\Im\lambda = 0$ and then no oscillatory component is present, corresponding to points in the extreme top-left region of the planes in Fig.~\ref{figure2}. For the set of larger networks, the dependence of $\langle \mathrm{INSI} \rangle $ and $\mathrm{CV}_{\mathrm{INSI}}$ on the $\Re\lambda{}$ is much sharper, as expected given the much smaller finite-size effects; this shrinks the available region, around the instability line, allowing for intermediate, more biologically plausible values of $\mathrm{CV}_{\mathrm{INSI}}$.

We remark that \NS{}s are highly non-linear and relatively stereotyped events, typical of an excitable non-linear system. The good predictive power of the linear analysis for the statistics of INSI signals that relatively small fluctuations around the system's fixed point, described well by a linear analysis, can ignite a \NS{}.

\subsection*{A spectrum of network events}
Our mean-field, finite-size network is a non-linear excitable system which, to the left of the Hopf bifurcation line, and close to it, can express different types of excursions from the otherwise stable fixed point. Large (almost stereotyped for high excitation) \NS{}s are exquisite manifestations of the non-linear excitable nature of the system, ignited by noise; the distribution of \NS{} size (number of spikes generated during the event) is relatively narrow and approximately symmetric (the Gaussian component of Eq.~\ref{eq.pSizeNSQO}).

\begin{figure}[h!]
\begin{center}
\setlength{\unitlength}{\textwidth}
\begin{picture}(1,0.8)

  \put(0.09,0.4)
  {
    \epsfig
    {
    file=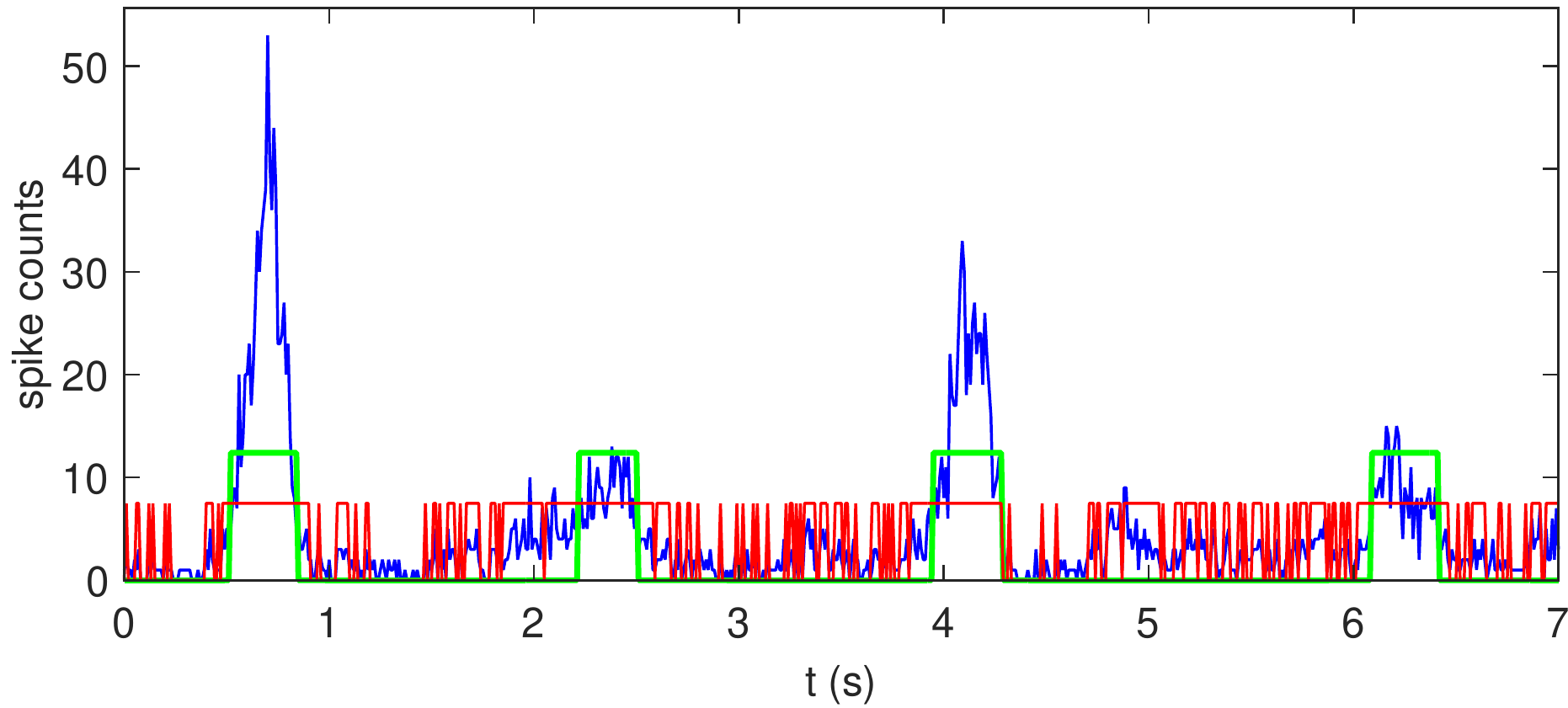,
    width=0.8\unitlength,
    }
  }
  \put(0.09,0)
  {
    \epsfig
    {
    file=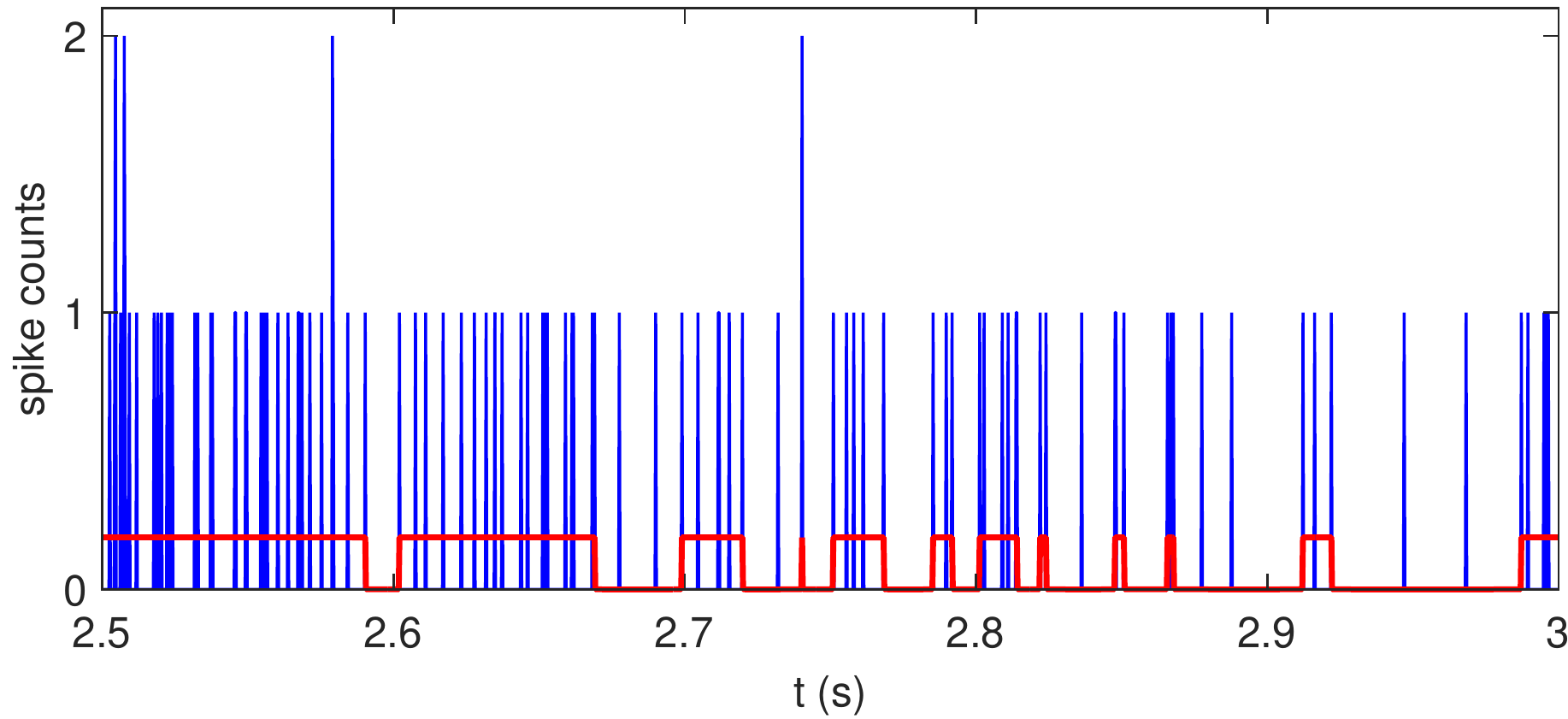,
    width=0.8\unitlength,
    }
  }
  \put(0.83,0.32){\makebox{\large \bf B}}
  \put(0.83,0.715){\makebox{\large \bf A}}
\end{picture}
\caption{{\bf Algorithms for network events detection.}
Panel A: total network activity from simulation (blue line) with detected \NS{}/quasi-orbits (green line) and avalanches (red line). Four large events (green line) are visible; the first and third are statistically classified as network spikes; the other smaller two as quasi-orbits. Note how network spikes and quasi-orbits are typically included inside a single avalanche. Panel B: a zoom over $0.5$ seconds of activity, with discretization time-step $0.25$ ms, illustrates avalanches structure (red line).}
\label{figure4}
\end{center}
\end{figure}

Noise can also induce smaller, transient excursions from the fixed point which can be adequately described as quasi-orbits in a linear approximation. In fact, noise induces a probability distribution on the size of such events, which can be computed as explained in Methods and Analysis (the exponential part in Eq.~\ref{eq.pSizeNSQO}). Fig.~\ref{figure4}, panel A, shows the activity of a simulated network (blue line) alongside with detected network events. We remark that the the different event types may not in general be easily distinguished on a single-event basis, while we argue that they are probabilistically distinguishable. From the best fit for the expected size distribution a threshold for the event size can be determined to separate events that are (\textit{a-posteriori}) more probably quasi-orbits from the ones that are more probably \NS{}s (for details, see Models and Analysis). Following such classification, the green line in Fig.~\ref{figure4}, panel A, marks the detection of two \NS{}s (first and third event) and two quasi-orbits (second and fourth event).

We also emphasize that the existence of quasi-orbits is a specific consequence of the fact that in the whole excitation-inhibition plane explored for the model, the low-activity fixed point becomes unstable via a Hopf bifurcation. It is indeed known that for nonlinear systems in the proximity of a Hopf bifurcation, noise promotes precursors of the bifurcation, which appear as transient synchronization events (see, \textit{e.g.}, \cite{khovanov2006spectral}). 

As one moves around the excitation-inhibition plane, to the left of the bifurcation line, the two types of events contribute differently to the overall distribution of network event sizes. Qualitatively, the farther from the bifurcation line, the higher the contribution of the small, ``quasi-linear'' events. This fact can be understood by noting that the average size of such events is expected to scale as $1 / |\Re \lambda|$, where $\Re \lambda$ is the real part of the dominant eigenvalue of the (stable) linearized dynamics (see Models and Analysis, Eq.~\ref{eq.Pssr2}). The average size is furthermore expected to scale with the amount of noise affecting the dynamics, thus the contribution of quasi-linear events is also expected to vanish for larger networks. 

It has been previously reported that activity dynamics may be different from one network to the other, reflecting idiosyncrasies of composition and history-dependent processes (\cite{van2005dynamics}). Moreover, the dynamics of a given network, as well as its individual neurons, may shift over time (minutes and hours) between different modes of activity (\cite{van2005dynamics,wagenaar2006persistent,gal2010dynamics}). We therefore chose to demonstrate the efficacy of our analytical approach on two data sets of large-scale random cortical networks.

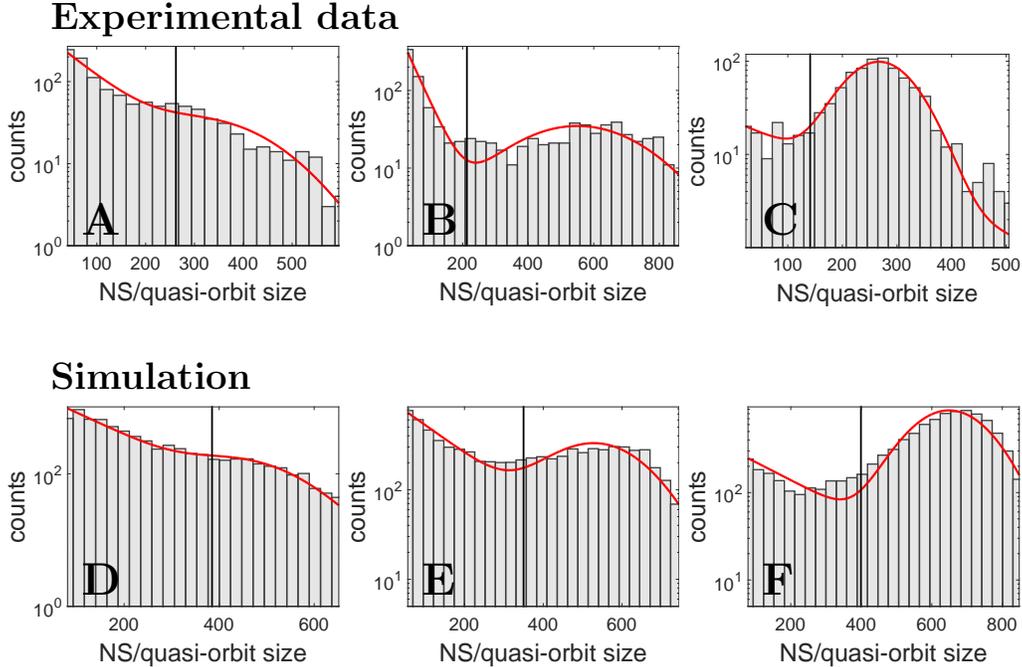
\begin{figure}[!htb]
\begin{center}
\setlength{\unitlength}{\textwidth}
\begin{picture}(1,0.65)

\put(-0.01,0.35)
  {
    \epsfig
    {
    file=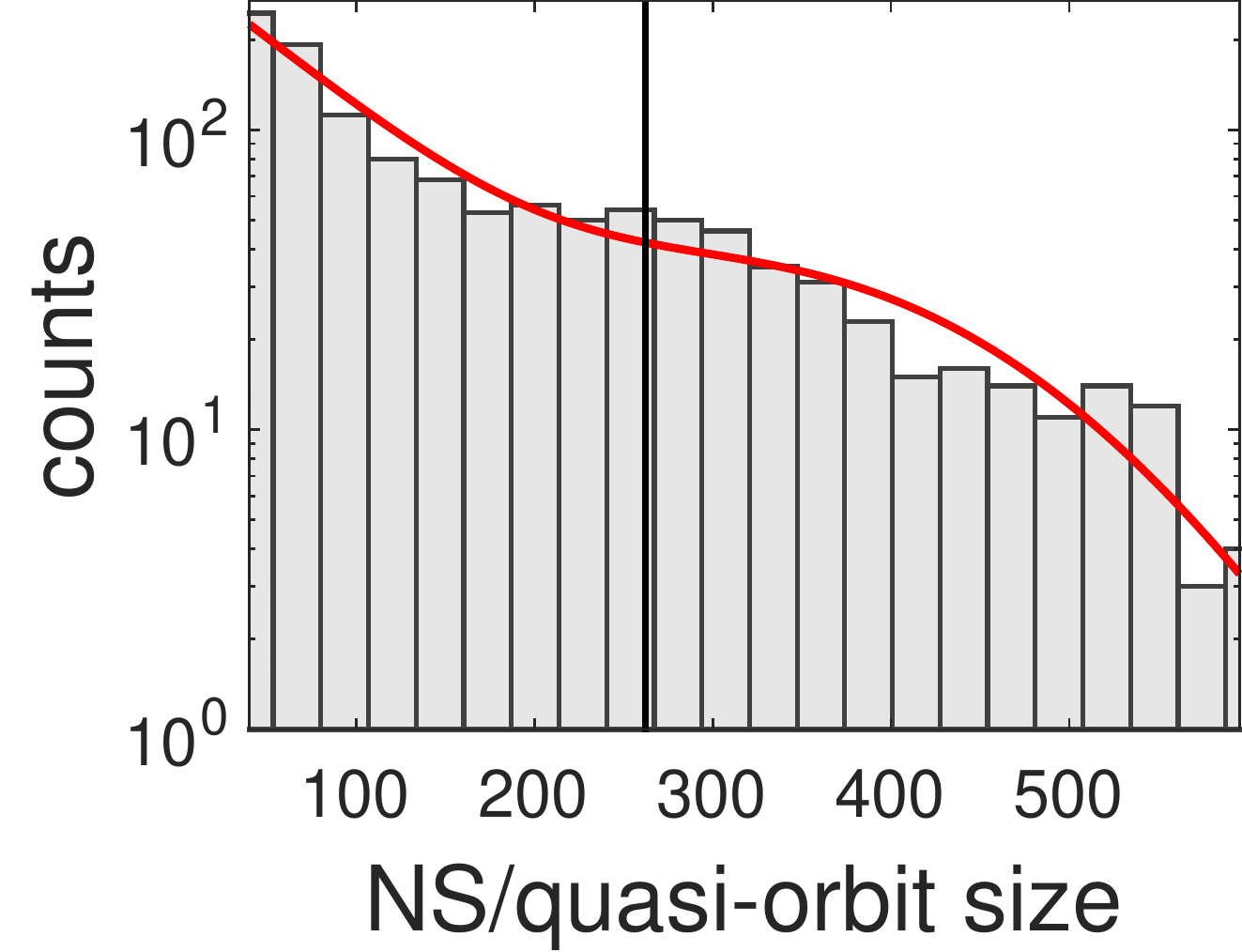,
    width=0.33\unitlength,
    }
  }
  \put(0.32,0.35)
  {
    \epsfig
    {
    file=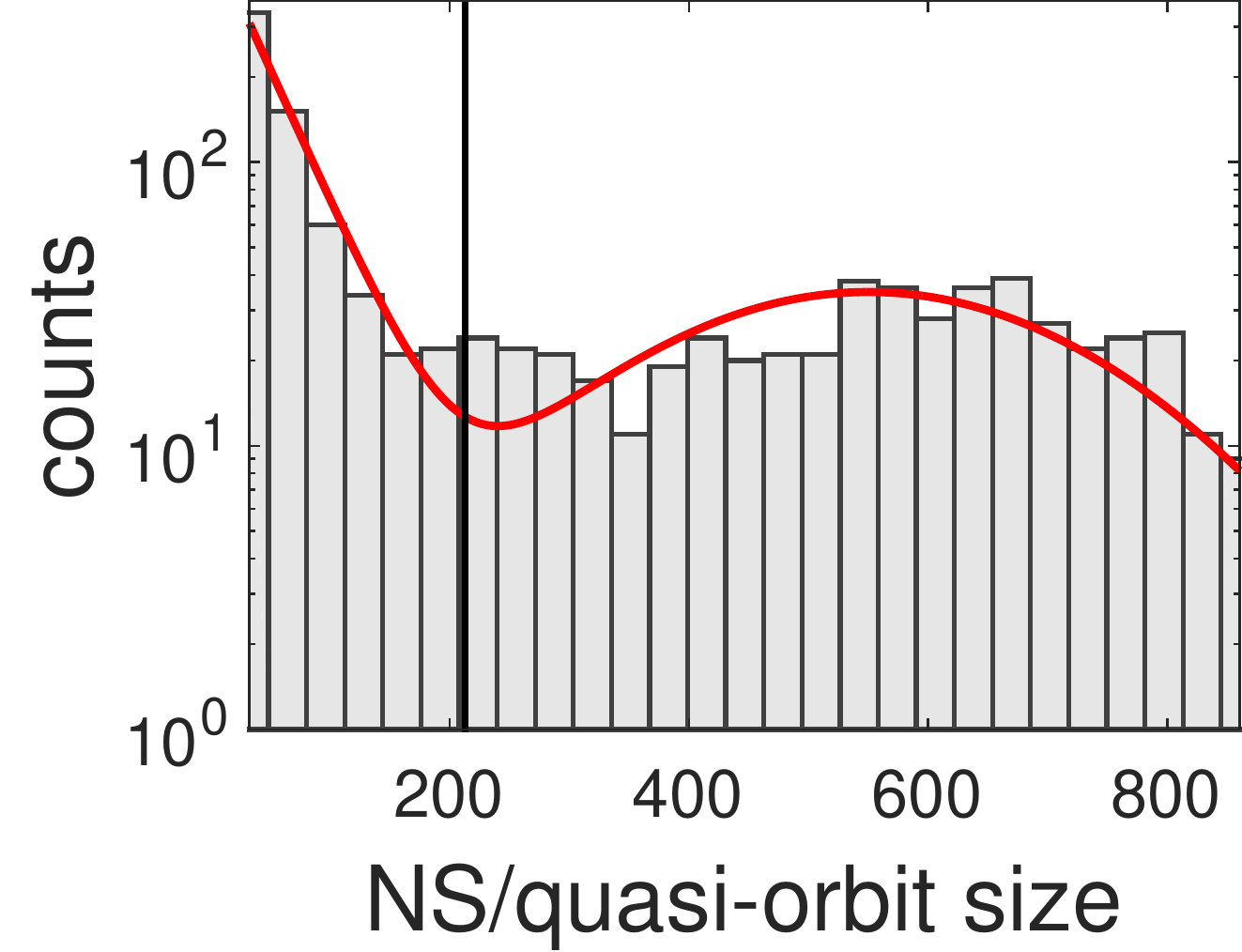,
    width=0.33\unitlength,
    }
  }
  \put(0.65,0.35)
  {
    \epsfig
    {
    file=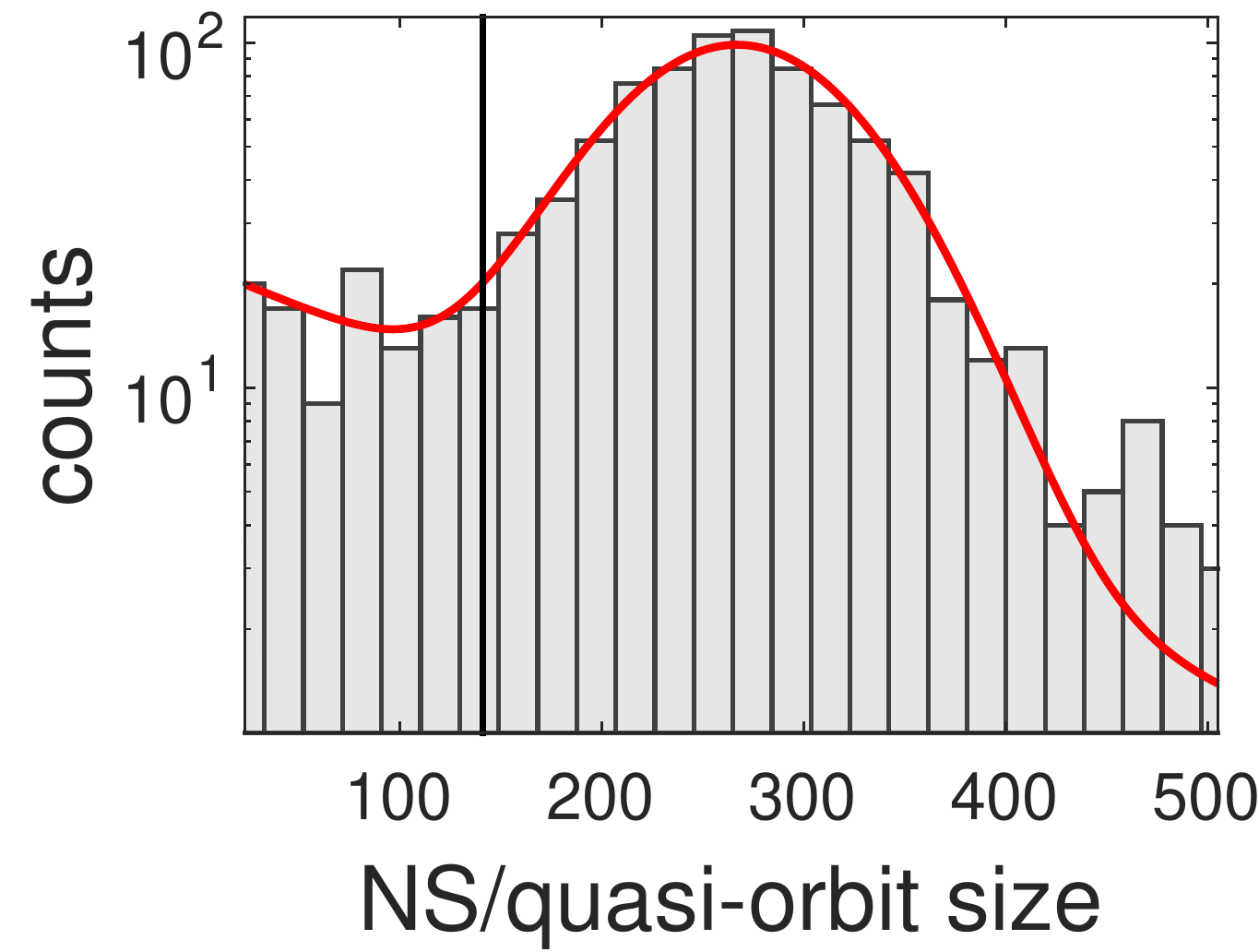,
    width=0.33\unitlength,
    }
  }
  
  \put(-0.01,0.)
  {
    \epsfig 
    {
    file=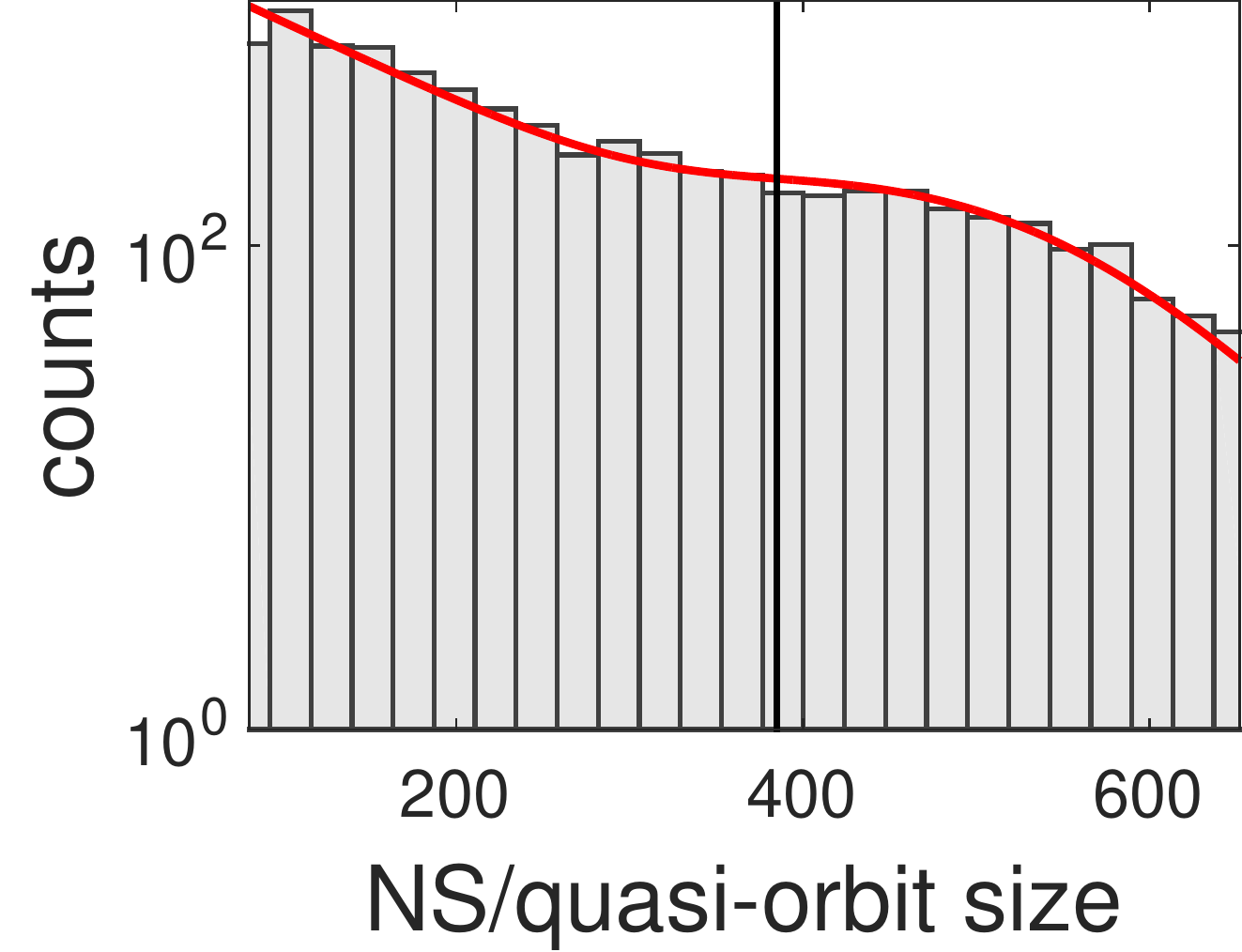,
    width=0.33\unitlength,
    }
  }
  \put(0.32,0.)
  {
    \epsfig
    {
    file=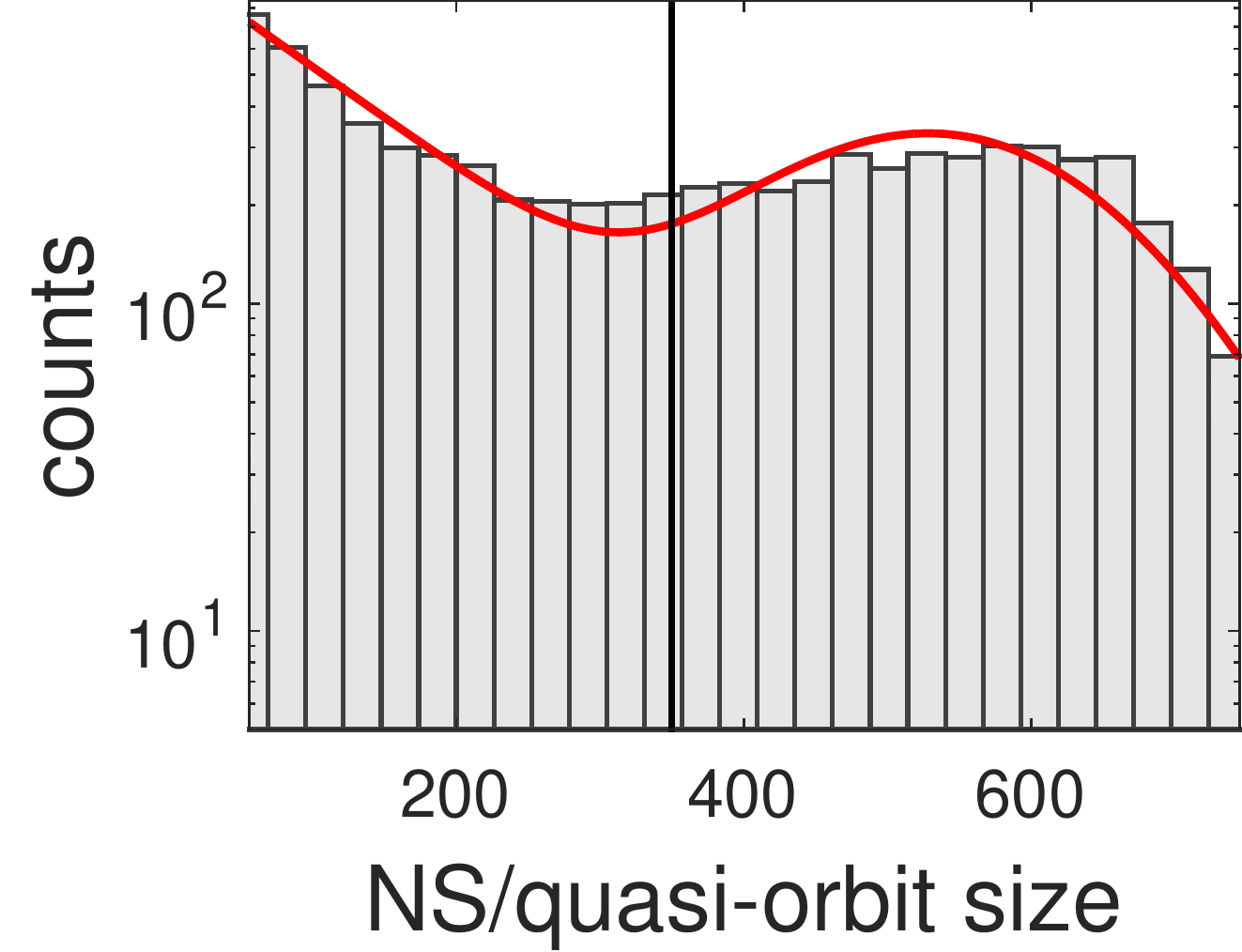,
    width=0.33\unitlength,
    }
  }
  \put(0.65,0.)
  {
    \epsfig
    {
    file=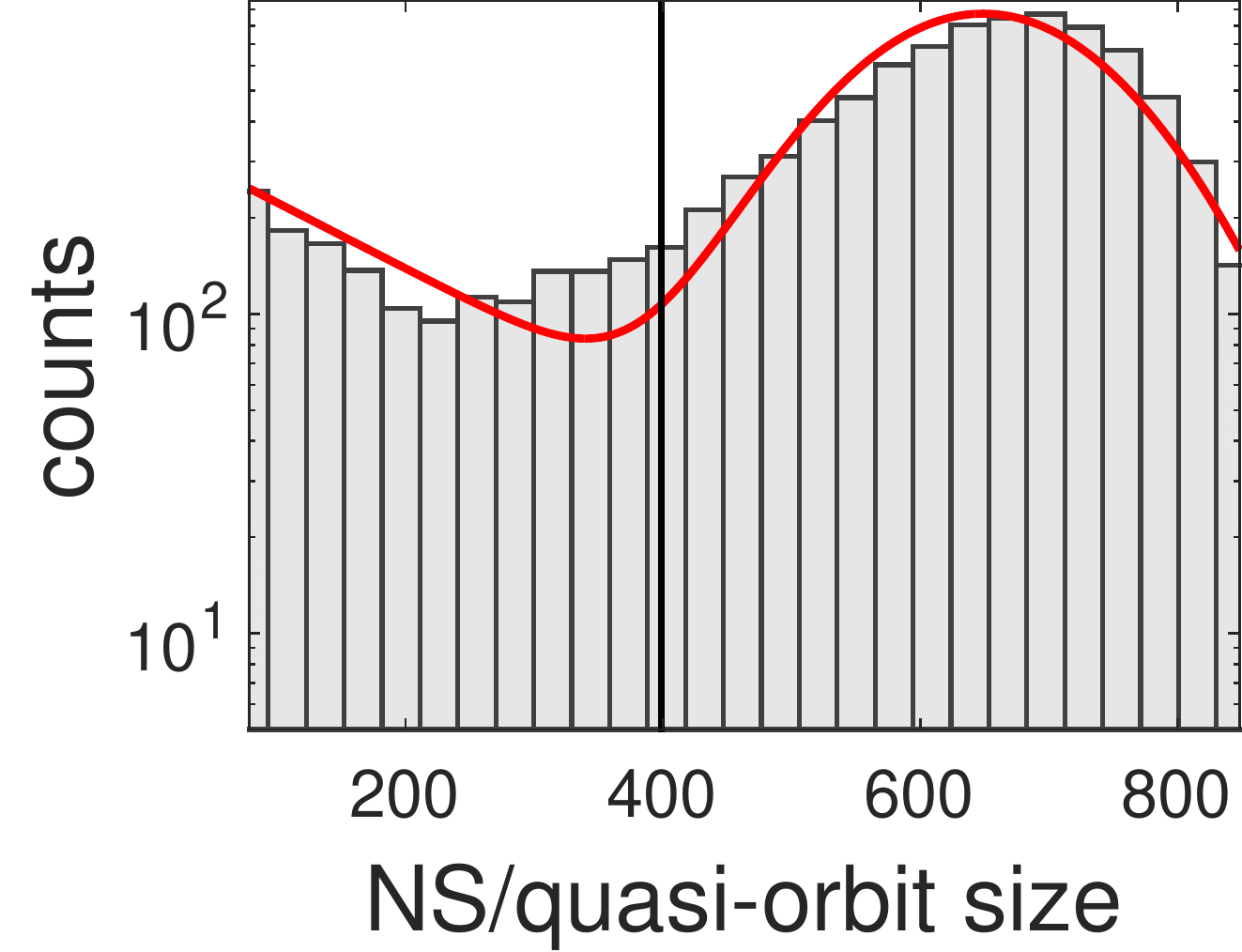,
    width=0.33\unitlength,
    }
  }
  \put(0.08,0.42){\makebox{\Large\textbf{A}}}
  \put(0.41,0.42){\makebox{\Large\textbf{B}}}
  \put(0.74,0.42){\makebox{\Large\textbf{C}}}
  \put(0.08,0.07){\makebox{\Large\textbf{D}}}
  \put(0.41,0.07){\makebox{\Large\textbf{E}}}
  \put(0.74,0.07){\makebox{\Large\textbf{F}}}
  
  \put(0.05,0.27){\makebox{\large \bf Simulation}}
  \put(0.05,0.62){\makebox{\large \bf Experimental data}}
\end{picture}
\caption{{\bf A broad spectrum of synchronous network events: simulations \textit{vs} \textit{ex-vivo} data.}
Panels A-C: experimental distributions of network events. Panels A and B: $\sim$40-minute recordings from a very long recording, for the same network; panel C: $\sim$1-hour recording from another cultured network. Panels D-F: distributions from simulations of networks corresponding to the points in Fig.~\ref{figure2} ($(\mathrm{w}_{\mathrm{exc}},\,\mathrm{w}_{\mathrm{inh}}) = (0.82, \, 0.7)$, $(\mathrm{w}_{\mathrm{exc}},\,\mathrm{w}_{\mathrm{inh}}) = (0.82, \, 0.55)$, $(\mathrm{w}_{\mathrm{exc}},\,\mathrm{w}_{\mathrm{inh}}) = (0.88, \, 0.55)$). The three networks of panels D-F have increasing levels of subcritical excitability. Note the logarithmic scale on the y-axis. The solid lines are fits of the theoretical distribution of event sizes, a sum of an exponential (for quasi-orbits) and a Gaussian (for \NS{}) distribution (see Models and Analysis, Eq.~\ref{eq.pSizeNSQO}). The vertical lines mark the probabilistic threshold separating \NS{} and quasi-orbits.}
\label{figure5}
\end{center}
\end{figure}

In panels A-C of Fig~\ref{figure5}, we show the experimental distributions of event sizes for two cultured networks: panels A and B are $\sim$40-minute recordings taken from a very long recording for the same network; panel C is $\sim$1-hour recording from a different cultured network. By visual inspection, the distributions appear to be consistent with two components contributing with various weights, both for different periods of the same network, and for different networks. In the light of the above theoretical considerations, one is led to generate the hypothesis that the two components contributing to the overall distribution were associated with quasi-orbits and network spikes respectively; to test this hypothesis, we fitted (solid lines in Fig.~\ref{figure5}) the experimental distributions with the sum of an exponential and a Gaussian distribution (see Models and Analysis, Eq.~\ref{eq.pSizeNSQO}), prepared to interpret a predominance of the exponential (Gaussian) component as a lesser (greater) excitability of the network. We remark that (see panels A and B) the relative weights of the two components appear to change over time for the same network, as if the excitability level would dynamically change; more on this at the end of this section.

To substantiate the above interpretation of experimental results, we turned to long simulations (about 5.5 hours) of networks in different points in the excitation-inhibition plane (Fig.~\ref{figure2}), from which we extracted the distribution of network events and fitted them with Eq.~\ref{eq.pSizeNSQO} as for experimental data (see panels D-F in Fig.~\ref{figure5}). Again, to the eye, the fits appear to be consistent with the two components variously contributing to the overall distribution, depending on the excitability of the network.

If, however, the fits are subject to a Kolmogorov-Smirnov test, the test fails ($p < 0.01$) for panels D and F. By inspecting the maximum distance between the cumulative distributions for simulation data and the fit, we found it at the lowest size bin for panel D, while the ``Gaussian'' part gives the greater mismatch for panel F. As for panel D, while the theoretical argument for the quasi-orbits clearly captures the shape of the size distributions, the way the test fails in the exponential part is interesting.

In fact, network events cannot be detected with arbitrarily small size: in a way, the detection procedure imposes a soft threshold on the event size, below which the exponential distribution is not applicable.

We can provide a rough estimate of such soft threshold as follows. A quasi-orbit duration is, to a first approximation, proportional to $1 / \Im \lambda$, which is of the order of few hundreds milliseconds not too far from the bifurcation line in the excitation-inhibition plane. Taking, for instance, 150 ms, an event will be detected if network activity within this time-span is larger than average (typically few spikes per second per neuron; we take 3 for the present example): this leads to a soft threshold of about 100 spikes. This would be the lower limit of applicability of the exponential part of the distribution; this also explains the trough observed for very small sizes.

As for the failure of the Kolmogorov-Smirnov test for the right part of the distribution in panel F, it should be remarked that the assumption of a Gaussian distribution for the size of network spikes, although generically plausible, is not grounded in a theoretical argument, and it's not surprising that, on the order of $10^4$ detected events, even a moderate skewness, as the one observed, can make the test fail.

The fit for experimental data of panels A-C passed the Kolmogorov-Smirnov test ($p > 0.01$).

As mentioned in the introduction, avalanches are cascades of neural activities clustered in time (see Models and Analysis for our operational definition; examples of different methods used in the literature to detect avalanches can be found in \cite{beggs2003neuronal,benayoun2010avalanches,poil2012critical,priesemann2014spike}). Fig.~\ref{figure4}, panel A and panel B, shows an example of the structure of the detected avalanches (red lines) in the model network.

We extracted avalanches from simulated data, as well as from experimental data.
For simulations, we choose data corresponding to three points in the $(\mathrm{w}_{\mathrm{exc}},\,\mathrm{w}_{\mathrm{inh}})$ plane of Fig.~\ref{figure2}, with constant $\mathrm{w}_{\mathrm{inh}} = 1$ and increasing $\mathrm{w}_{\mathrm{exc}}$, with the rightmost falling exactly over the instability line (white solid line in Fig.~\ref{figure2}). Three experimental data sets were extracted from different periods of a very long recording of spontaneous activity from a neural culture; each data set is a $40$-minute recording.

\begin{figure}[!htb]
\begin{center}
\setlength{\unitlength}{\textwidth}
\begin{picture}(1,0.65)

  \put(-0.01,0.35)
  {
    \epsfig
    {
    file=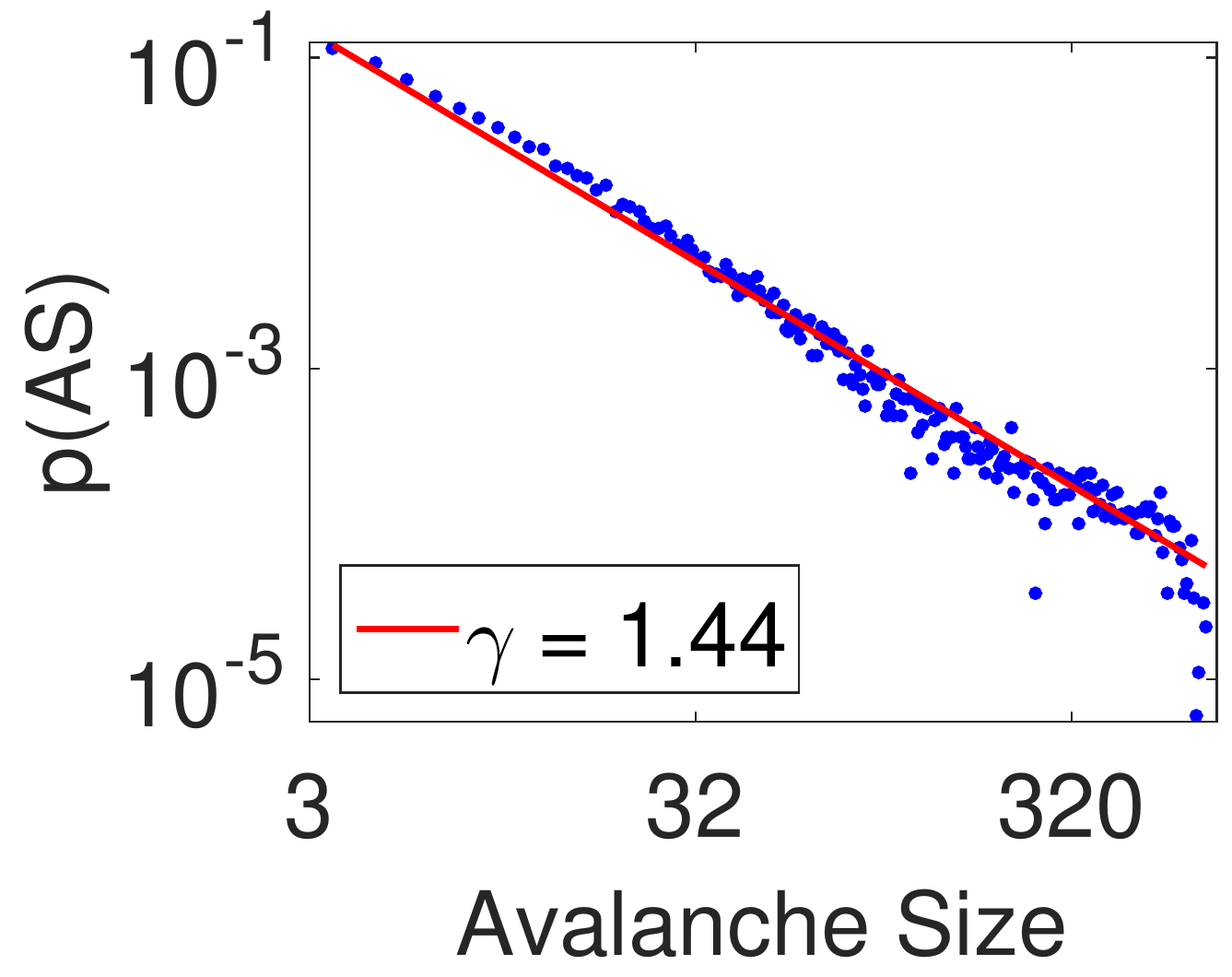,
    width=0.33\unitlength,
    }
  }
  \put(0.32,0.35)
  {
    \epsfig
    {
    file=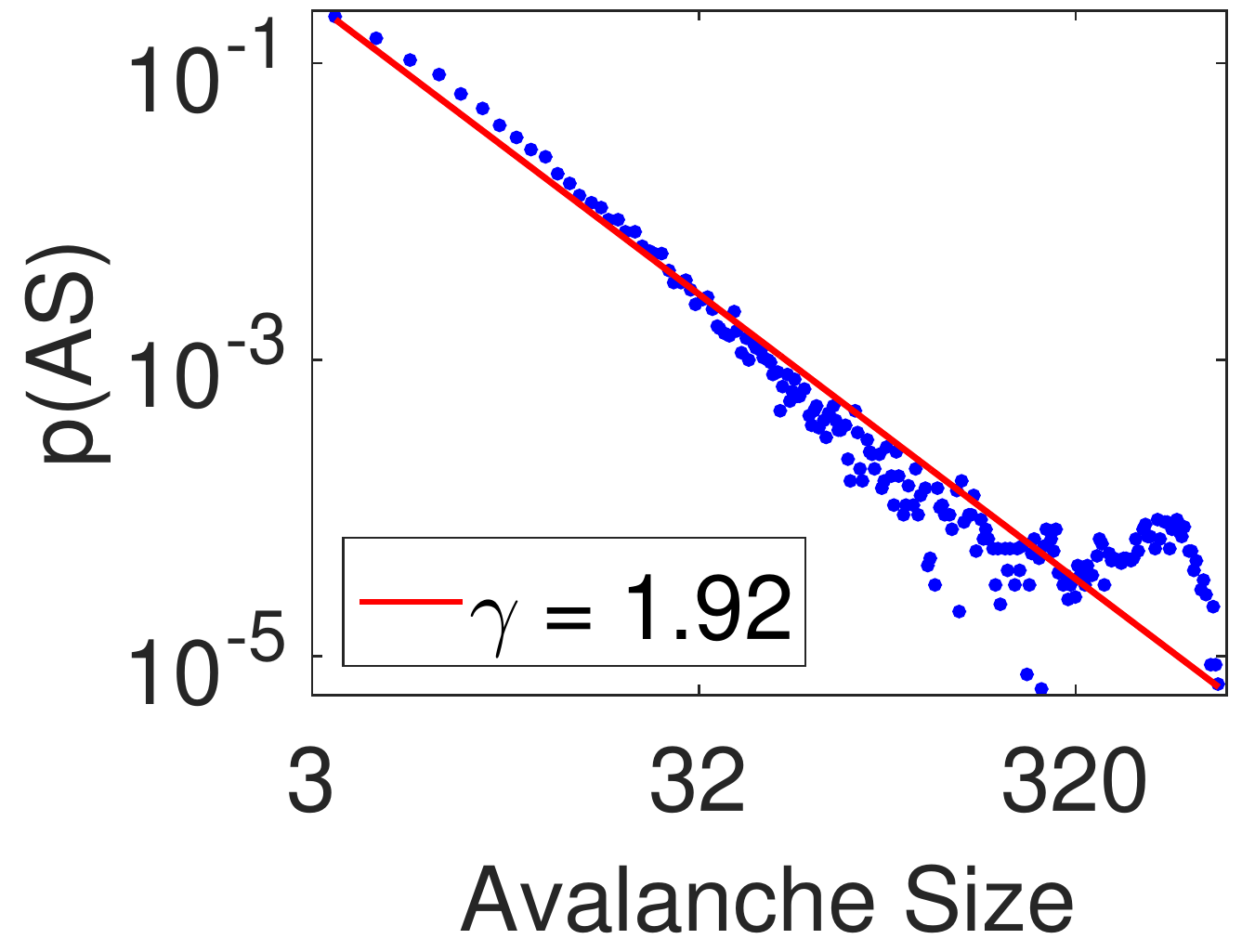,
    width=0.33\unitlength,
    }
  }
  \put(0.65,0.35)
  {
    \epsfig
    {
    file=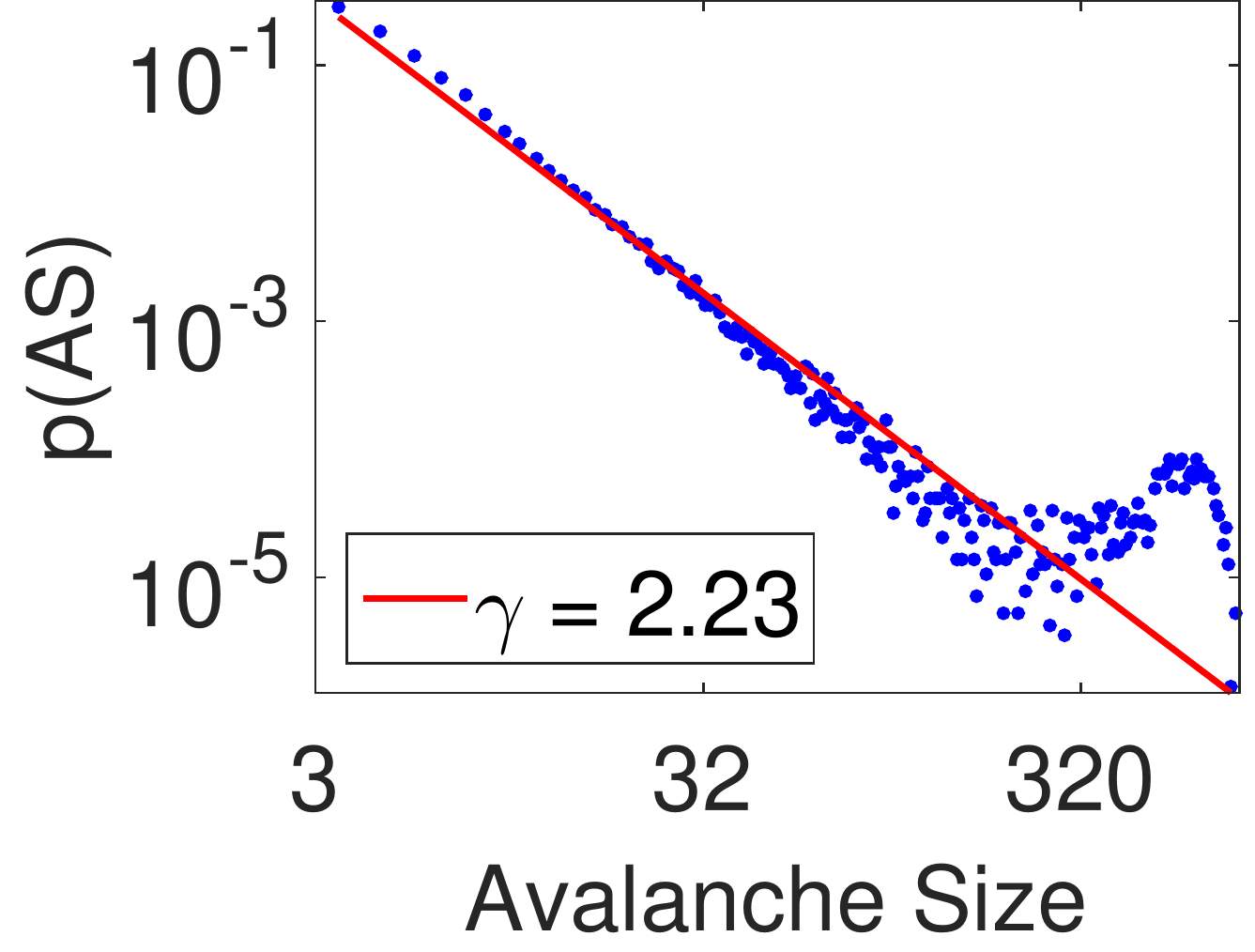,
    width=0.33\unitlength,
    }
  }
  
  \put(-0.01,0.)
  {
    \epsfig 
    {
    file=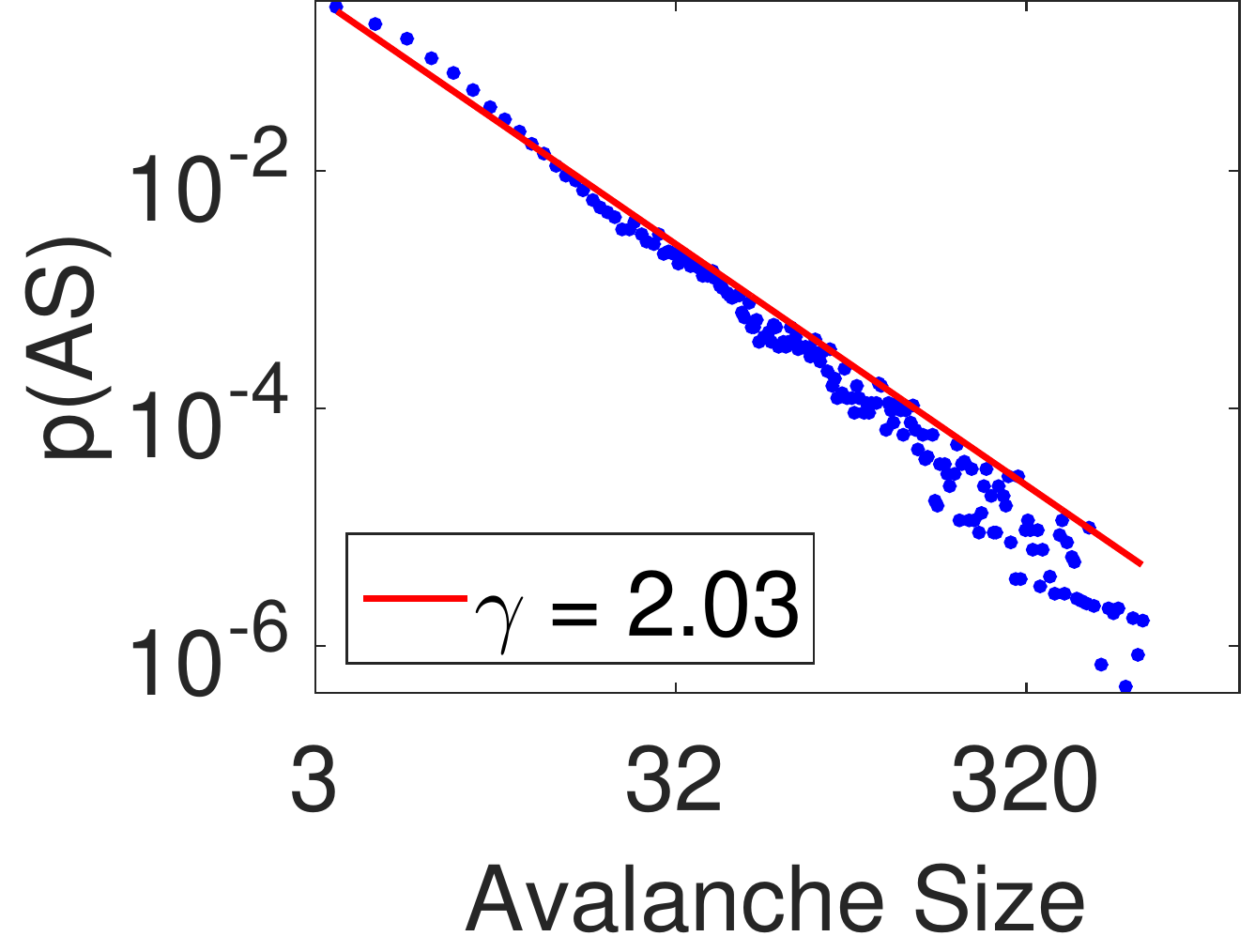,
    width=0.33\unitlength,
    }
  }
  \put(0.32,0.)
  {
    \epsfig
    {
    file=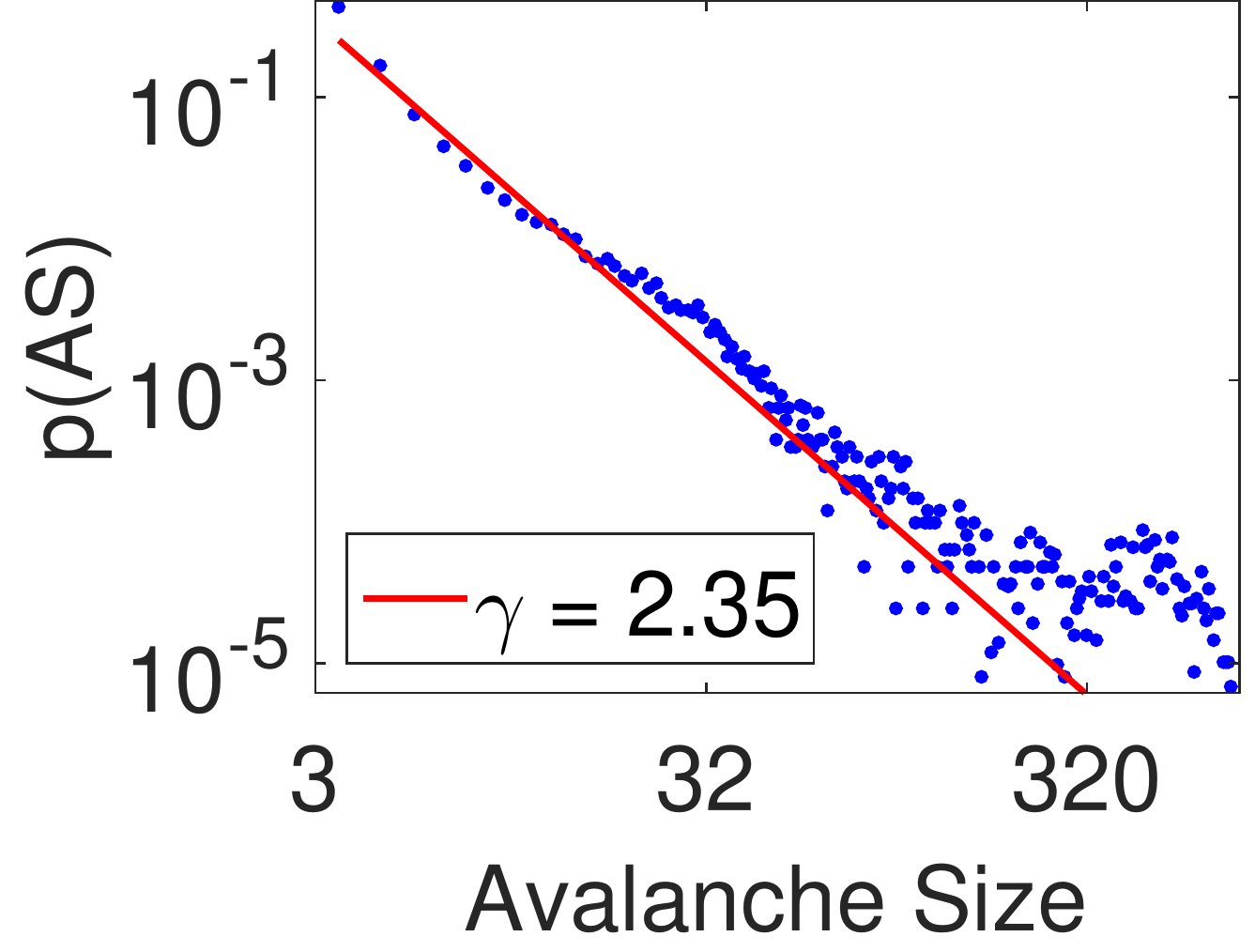,
    width=0.33\unitlength,
    }
  }
  \put(0.65,0.)
  {
    \epsfig
    {
    file=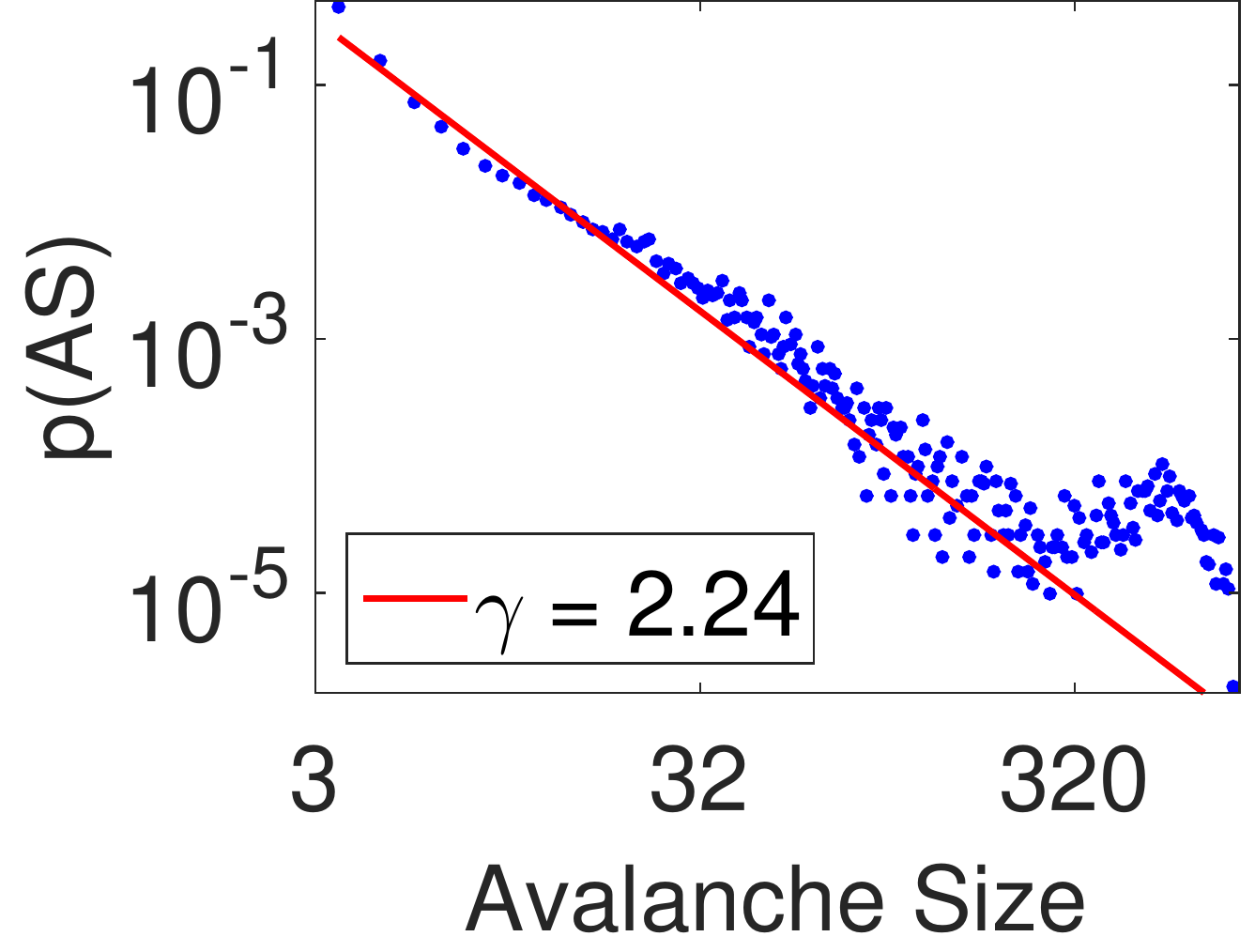,
    width=0.33\unitlength,
    }
  }
  \put(0.08,0.4){\makebox{\Large\textbf{A}}}
  \put(0.41,0.4){\makebox{\Large\textbf{B}}}
  \put(0.74,0.4){\makebox{\Large\textbf{C}}}
  \put(0.08,0.05){\makebox{\Large\textbf{D}}}
  \put(0.41,0.05){\makebox{\Large\textbf{E}}}
  \put(0.74,0.05){\makebox{\Large\textbf{F}}}
  
  \put(0.05,0.62){\makebox{\large \bf Simulation}}
  \put(0.05,0.27){\makebox{\large \bf Experimental data}}
\end{picture}
\caption{{\bf Avalanche size distribution: simulations \textit{vs} \textit{ex-vivo} data.}
Panels A-C: mean-field simulations, with fixed inhibition $\mathrm{w}_{\mathrm{inh}} = 1.$ and increasing excitation ($\mathrm{w}_{\mathrm{exc}} = 0.9, \, 0.94, \, 1$). The distributions are well fitted by power-laws; panel B and C clearly show the buildup of `bumps' in the high-size tails, reflecting the increasing contribution from network spikes and quasi-orbits in that region of the distribution. Panels D-F from \textit{ex-vivo} data, different $\sim 40$-minute segments from one long recording; power-laws are again observed, although fitted exponents cover a smaller range; in panels $E$ and $F$, bumps are visible, similar to model findings. The similarity between the theoretical and experimental distributions could reflect changes of excitatory/inhibitory balance in time in the experimental preparation. Since all the three simulations lay on the left of or just on the bifurcation line (white line in Fig.~\ref{figure2}), the shown results are compatible with the experimental network operating in a slightly sub-critical regime.}
\label{figure6}
\end{center}
\end{figure}

In Fig.~\ref{figure6} we show (in log-log scale) the distribution of avalanche sizes for the three simulated networks (top row) and the three experimental (bottom row) data sets (blue dots); red lines are power-law fits \cite{clauset2009power}.

From the panels in the top row we see that the distributions are well fitted, over a range of two orders of magnitude, by power-laws with exponents ranging from about $1.5$ to about $2.2$, consistent with the results found in \cite{beggs2003neuronal}. Note that in the cited paper the algorithm used for avalanche detection is quite different from ours, and the wide range of power-law exponents is related to their dependence on the time-window used to discretize data. In \cite{mazzoni2007dynamics} (adopting yet another algorithm for avalanche detection), both the shape of the avalanche distribution and the exponent vary depending on using pharmacology to manipulate synaptic transmission, over a range compatible with our model findings; notably, they find the slope of the power-law to be increasing with the excitability of the network, which is consistent with our modeling results.

Panels B and C of Fig.~\ref{figure6} clearly show the buildup of `bumps' in the high-size tails, increasing with the self-excitation of the network; this is understood as reflecting the predominance of a contribution from \NS{} and possibly quasi-orbits in that region of the distribution, on top of a persisting wide spectrum of avalanches. This feature also is consistent with the experimental findings of \cite{mazzoni2007dynamics}, and has been previously shown in a theoretical model \cite{levina2009phase} for non-leaky integrate-and-fire neurons endowed with \STD{} and synaptic facilitation.

Turning to the plots in the bottom row of Fig.~\ref{figure6}, we observe the following features: power-laws are again observed over two decades and more; in panels $E$ and $F$, bumps are visible, similar to model findings; power-law exponents cover a smaller range just above 2.

While the sequence of plots in two rows (modeling and experiment) clearly shows similar features, we emphasize that experimental data were extracted from a unique long recording, with no intervening pharmacological manipulations affecting synaptic transmission; on the other hand, it has been suggested \cite{haroush2015slow} that a dynamic modulation of the excitatory/inhibitory balance can indeed be observed in long recordings; although our model would be inherently unable to capture  such effects, it is tempting to interpret the suggestive similarity between the theoretical and experimental distributions in Fig.~\ref{figure6} as a manifestation of such changes of excitatory/inhibitory balance in time, of which the theoretical distributions would be a `static' analog. To rule out the possibility that different behaviors in time could be due to intrinsic and global modifications in the experimental preparation, we checked (see Fig.~\ref{figureS1} - S1 Fig) the waveforms of the recorded spikes across all MEA electrodes, comparing the earliest and latest used recordings (about 40 minutes each, separated by about 34 hours). In most cases the waveforms for the two recordings are remarkably similar, and when they are not, no systematic trend in the differences is observed.

If our interpretation is correct, the experimental preparation operates below, and close, to an oscillatory instability; on the other hand, contrary to \NS{}, the appearance of avalanches does not seem to be exquisitely related to a Hopf bifurcation, rather they seem to generically reflect the non-linear amplification of spontaneous fluctuations around an almost unstable fixed point -- a related point will be mentioned in the next section. We also remark that we obtain power-law distributed avalanches in a (noisy) mean-field rate model, by definition lacking any spatial structure; while the latter could well determine specific (possibly repeating) patterns of activations (as observed in \cite{beggs2004neuronal}), it is here suggested to be not necessary for power-law distributed avalanches.

The avalanche size distribution for the same network as in Fig.~\ref{figure5}, panel C, is sparser but qualitatively compatible with the distribution in Fig.~\ref{figure6}, panel F (see Fig.~\ref{figureS2} - S2 Fig); in particular, the distribution shows a prominent peak for high-size avalanches, consistently with the interpretation, given in connection with Fig.~\ref{figure5}, of high excitability.

We do not provide examples of avalanche and \NS{}-quasi orbits size distributions in the super-critical region on the right of the Hopf bifurcation line in Fig.~\ref{figure2}; this is because the phenomenology in that region is relatively stereotyped and easy to guess/understand: the high excitability of the network generates, moving on the right of the bifurcation line, increasingly stereotyped network spikes, which dominate the size distribution of the network events (see Fig.~\ref{figureS3} - S3 Fig, panel A); even though finite-size fluctuations blur the bifurcation line, quasi-orbits are expected to contribute very little in the supercritical region; the distribution of avalanche sizes is increasingly dominated by the high-size bump associated with network spikes (see Fig.~\ref{figureS3} - S3 Fig, panel B).

\subsection*{Inferring the time-scales}
The fatigue mechanism at work (\STD{} in our case) is a key element of the transient network events, in its interplay with the excitability of the system. While the latter can be manipulated through pharmacology, \STD{} itself (or spike frequency adaptation, another neural fatigue mechanism) cannot be directly modulated. It is therefore interesting to explore ways to infer relevant properties of such fatigue mechanisms from the experimentally accessible information, i.e. the firing activity of the network. We focus in the following on deriving the effective (activity-dependent) time scale  of \STD{} from the sampled firing history.

The starting point is the expectation that the fatigue level just before a \NS{} should affect the strength of the subsequent \NS{}. We therefore measured the correlation between $r$ (fraction of available synaptic resources) and the total number of spikes emitted during the \NS{} (\NS{} size) from simulations. We found that the average value of $r$ just before a \NS{} is an effective predictor of the \NS{} size, the more so as the excitability of the network grows.

Based on the $r$-\NS{} size correlation, we took the above ``experimental'' point of view, that only the firing activity $\nu$ is directly observable, while $r$ is not experimentally accessible. Furthermore, the success of the linear analysis for the inter-\NS{} interval statistics (due to the \NS{} being a low-threshold very non-linear phenomenon), suggests that without assuming a specific form for the dynamics of the fatigue variable $f$, we may tentatively adopt for it a generic linear integrator form, of which we want to infer the characteristic time-scale $\tau^*$:
\begin{equation}
\label{eq.linearIntegrator}
\dot{f} = -\frac{f}{\tau^*} + \nu(t)
\end{equation} 

\begin{figure}[!htb]
\begin{center}
\setlength{\unitlength}{\textwidth}
\begin{picture}(1,0.35)

  \put(-0.01,0)
  {
    \epsfig
    {
    file=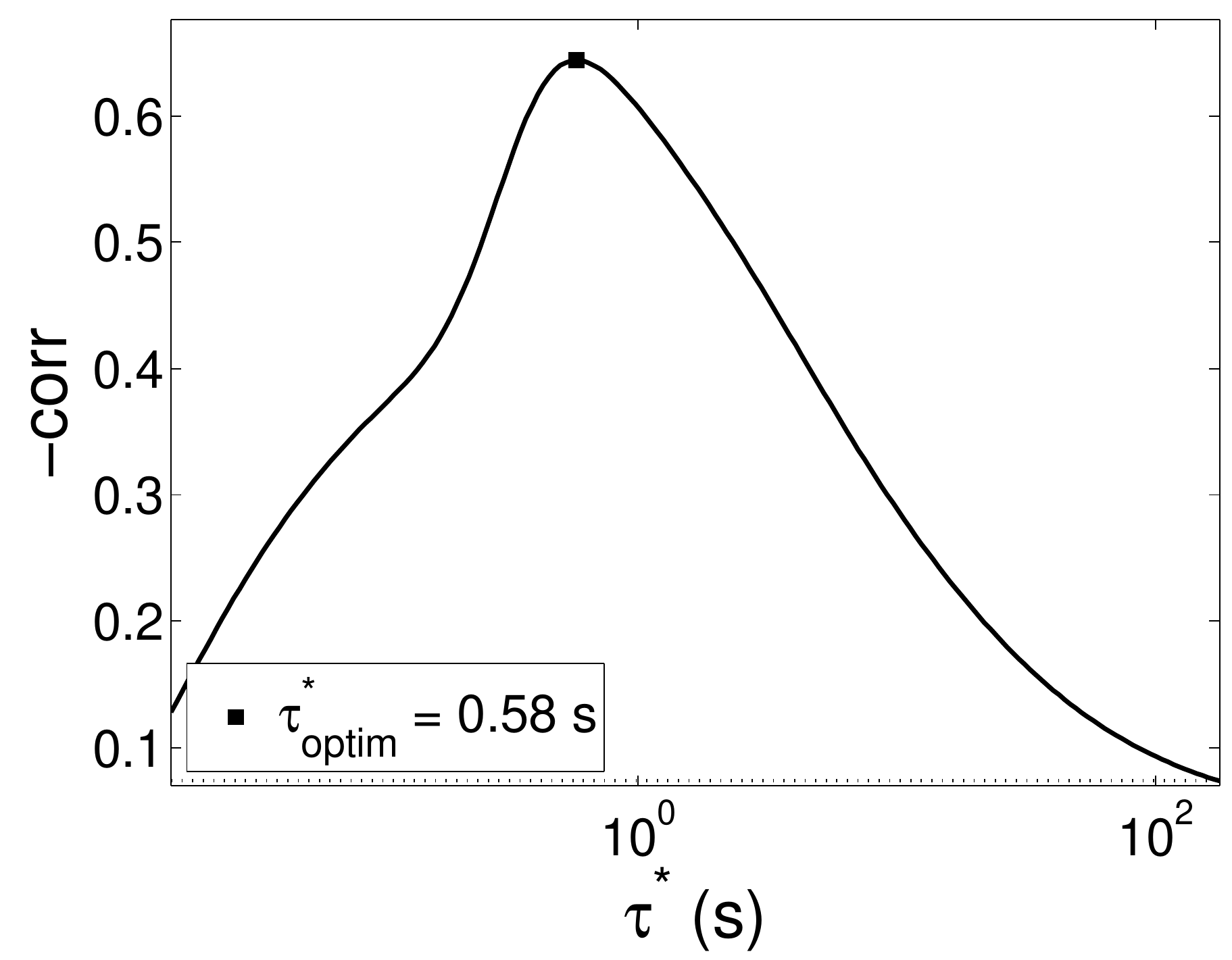,
    width=0.45\unitlength,
    }
  }
  \put(0.5,0)
  {
    \epsfig
    {
    file=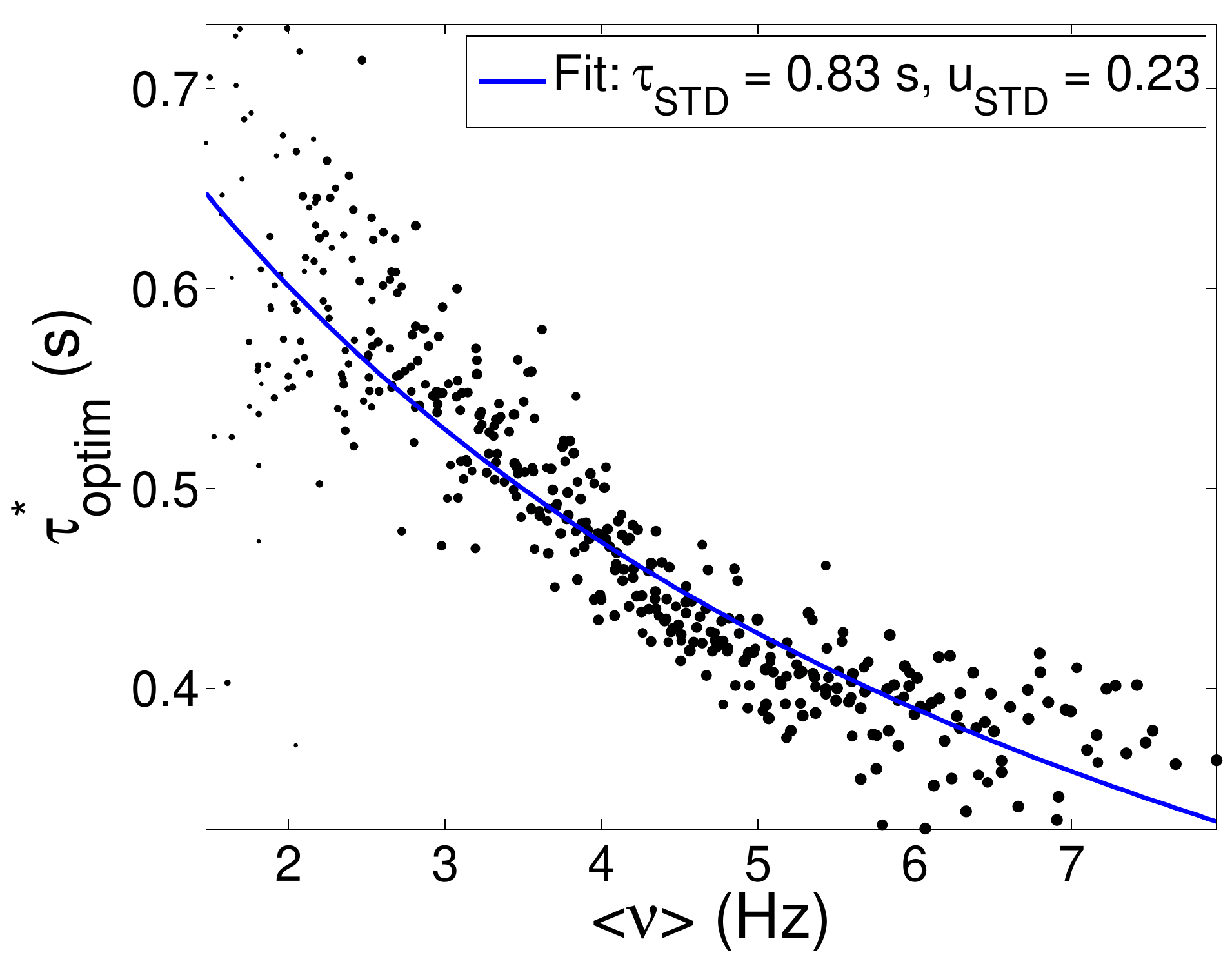,
    width=0.45\unitlength,
    }
  }
  \put(0.075,0.3){\makebox{\Large\textbf{A}}}
  \put(0.595,0.3){\makebox{\Large\textbf{B}}}
\end{picture}
\caption{{\bf Slow time-scales inference procedure: test on simulation data.}
Panel A: correlation between low-pass filtered network activity $f$ (see Eq.~\ref{eq.linearIntegrator}) and the size of the immediately subsequent network spike plotted against the time-scale $\tau^*$ of the low-pass integrator (continuous line). The correlation presents a clear (negative) peak for an `optimal' value $\tau^*_{\mathrm{optim}} = 0.58$ s of the low-pass integrator; such value is interpreted as the effective time-scale of the putative slow self-inhibitory mechanism underlying the statistics of network events -- in this case, short-term synaptic depression (\STD{}); as a reference, the dotted line marks the value computed for surrogate data (see text). Panel B: for each point in the $(\mathrm{w}_{\mathrm{exc}}, \, \mathrm{w}_{\mathrm{inh}})$-plane (see Fig.~\ref{figure2}), $\tau^*_{\mathrm{optim}}$ \textit{vs} average network activity; the continuous line is the best fit of the theoretical expectation for \STD{}'s effective time-scale (Eq.~\ref{eq.tauStarR}); the fitted values for the \STD{} parameters $\tau_{\mathrm{STD}}$ and $u_{\mathrm{STD}}$ are consistent with the actual values used in simulation ($\tau_{\mathrm{STD}} = 0.8$ s, $u_{\mathrm{STD}} = $ 0.2).}
\label{figure7}
\end{center}
\end{figure}

To do this, first we reconstruct $f(t)$ from $\nu(t)$ for a given $\tau^*$; then we set up an optimization procedure to estimate $\tau^*_{\mathrm{optim}}$, based on the maximization of the (negative) $f$-\NS{} size correlation (a strategy inspired by a similar principle was adopted in \cite{linaro2011inferring}). Fig.~\ref{figure7}, panel A, shows an illustrative example of how the correlation peaks around the optimal value. As a reference, the dotted line marks the value below which $95\%$ of the correlations computed from surrogate data fall; surrogate data are obtained by shuffling the values of $f$ at the beginning of each \NS{}.

We remark that in this analysis we use both \NS{} and quasi-orbit events (which are both related to the proximity to a Hopf bifurcation). This is reasonable since we expect to gain more information about the anti-correlation between $f$ and \NS{} size by including both types of large network events.

Although the procedure successfully recovers a maximum in the correlation, the value of $\tau^*_{\mathrm{optim}}$ (0.58 s) reported in Fig.~\ref{figure7}, panel A, is not close to the value of $\tau_{\mathrm{STD}}$ (0.8 s). Yet this is expected, since in Eq.~\ref{eq.linearIntegrator}, $\tau^*$ will in general depend on $\tau_{\mathrm{STD}}$ and other parameters of the dynamics, but also on the point around which the dynamics is being linearized, more precisely on the average activity $\langle \nu \rangle$. Specifically, when the fatigue variable follows the Tsodyks-Markram model of \STD{} (which of course was actually the case in the simulations), linearizing the dynamics of $r$ around a fixed point $\langle r \rangle$ ($\langle r \rangle = 1 / (1 + u_{\mathrm{STD}}\,\langle \nu \rangle\,\tau_{\mathrm{STD}})$), $r$ behaves as a simple linear integrator with a time-constant:
\begin{equation}\label{eq.tauStarR}
\tau^*_{\mathrm{optim}} = \tau_{\mathrm{STD}} \, \langle r \rangle = \frac{\tau_{\mathrm{STD}}}{1 + u_{\mathrm{STD}}\,\langle \nu \rangle \,\tau_{\mathrm{STD}}}
\end{equation}
that depends on $\tau_{\mathrm{STD}}$, $u_{\mathrm{STD}}$, and $\langle \nu \rangle$.

To test this relationship, we performed the optimization procedure for each point of the excitation-inhibition plane. The optimal $\tau^*$ values across the excitation-inhibition plane against $\langle \nu \rangle$ are plotted in Fig.~\ref{figure7}, panel B (dots). The solid line is the best fit of $\tau_{\mathrm{STD}}$ and $u_{\mathrm{STD}}$ from Eq.~\ref{eq.tauStarR}, which are consistent with the actual values used in simulations.

This result is suggestive of the possibility of estimating from experiments the time-scale of an otherwise inaccessible fatigue variable, by modeling it as a generic linear integrator, with a ``state dependent'' time-constant.

\begin{figure}[!htb]
\begin{center}
\setlength{\unitlength}{\textwidth}
\begin{picture}(1,0.35)

  \put(-0.01,0)
  {
    \epsfig
    {
    file=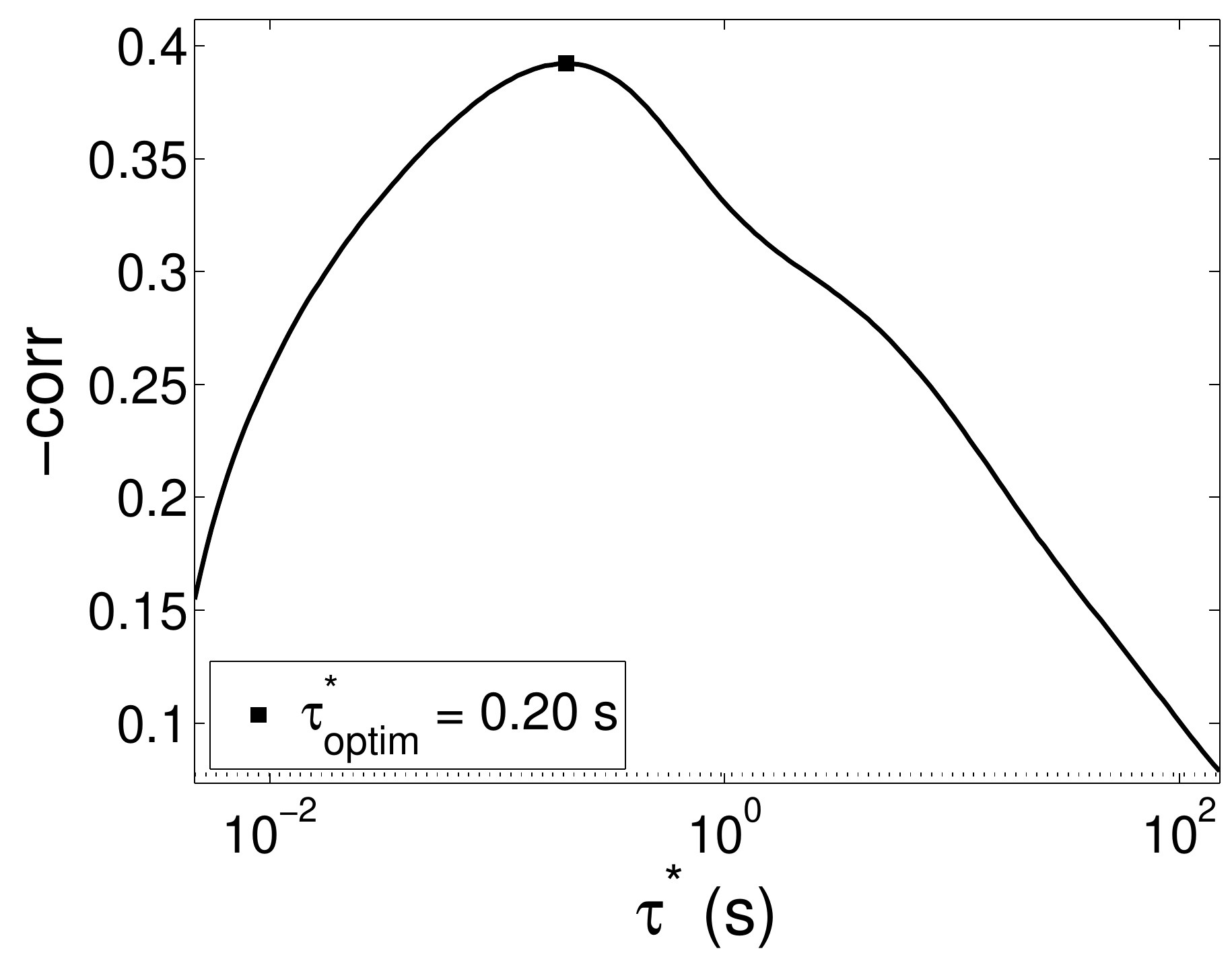,
    width=0.45\unitlength,
    }
  }
  \put(0.5,0)
  {
    \epsfig
    {
    file=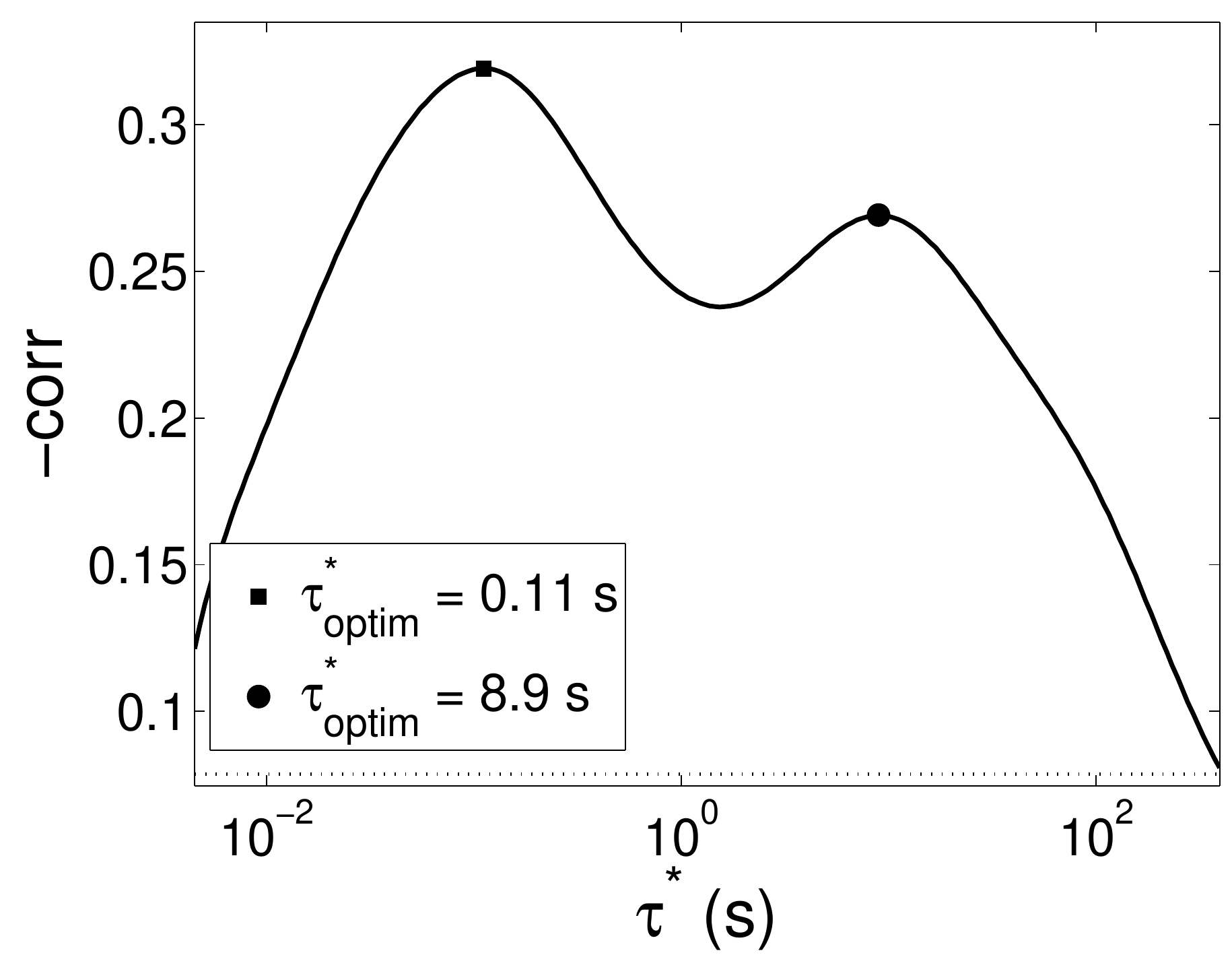,
    width=0.45\unitlength,
    }
  }
  \put(0.085,0.3){\makebox{\Large\textbf{A}}}
  \put(0.595,0.3){\makebox{\Large\textbf{B}}}
\end{picture}
\caption{{\bf Slow time-scales inference procedure on \textit{ex-vivo} data.}
Correlation between low-pass filtered network activity $f$ (see Eq.~\ref{eq.linearIntegrator}) and the size of the immediately subsequent network spike plotted against the time-scale $\tau^*$ of the low-pass integrator for two experimental datasets (different periods -- about $40$ minutes each -- in a long recording). The plot in panel A is qualitatively similar to the simulation result shown in panel A of Fig.~\ref{figure7}: a peak, although broader and of smaller maximum (absolute) value, is clearly identified and statistically significant (with respect to surrogate data, dotted line). Panel B shows two significant peaks in the correlation plot, a possible signature of two concurrently active fatigue processes, with time scales differing by roughly an order of magnitude. Panel A: same data as Fig.~\ref{figure5}, panel B.}
\label{figure8}
\end{center}
\end{figure}

Fig.~\ref{figure8} shows the outcome of the same inference procedure for two segments of experimental recordings. The plot in panel A is qualitatively similar to panel A in Fig.~\ref{figure7}: although the peak is broader and the maximum correlation (in absolute value) is smaller, the $\tau^*$ peak is clearly identified and statistically significant (with respect to surrogates, dotted line), thus suggesting a dominant time scale for the putative underlying, unobserved fatigue process. However, Fig.~\ref{figure8}, panel B, clearly shows two significant peaks in the correlation plot; it would be natural to interpret this as two fatigue processes, with time scales differing by an order of magnitude, simultaneously active in the considered recording segment.

\begin{figure}[!htb]
\begin{center}
\setlength{\unitlength}{\textwidth}
\begin{picture}(1,0.35)
  \put(-0.01,0)
  {
    \epsfig
    {
    file=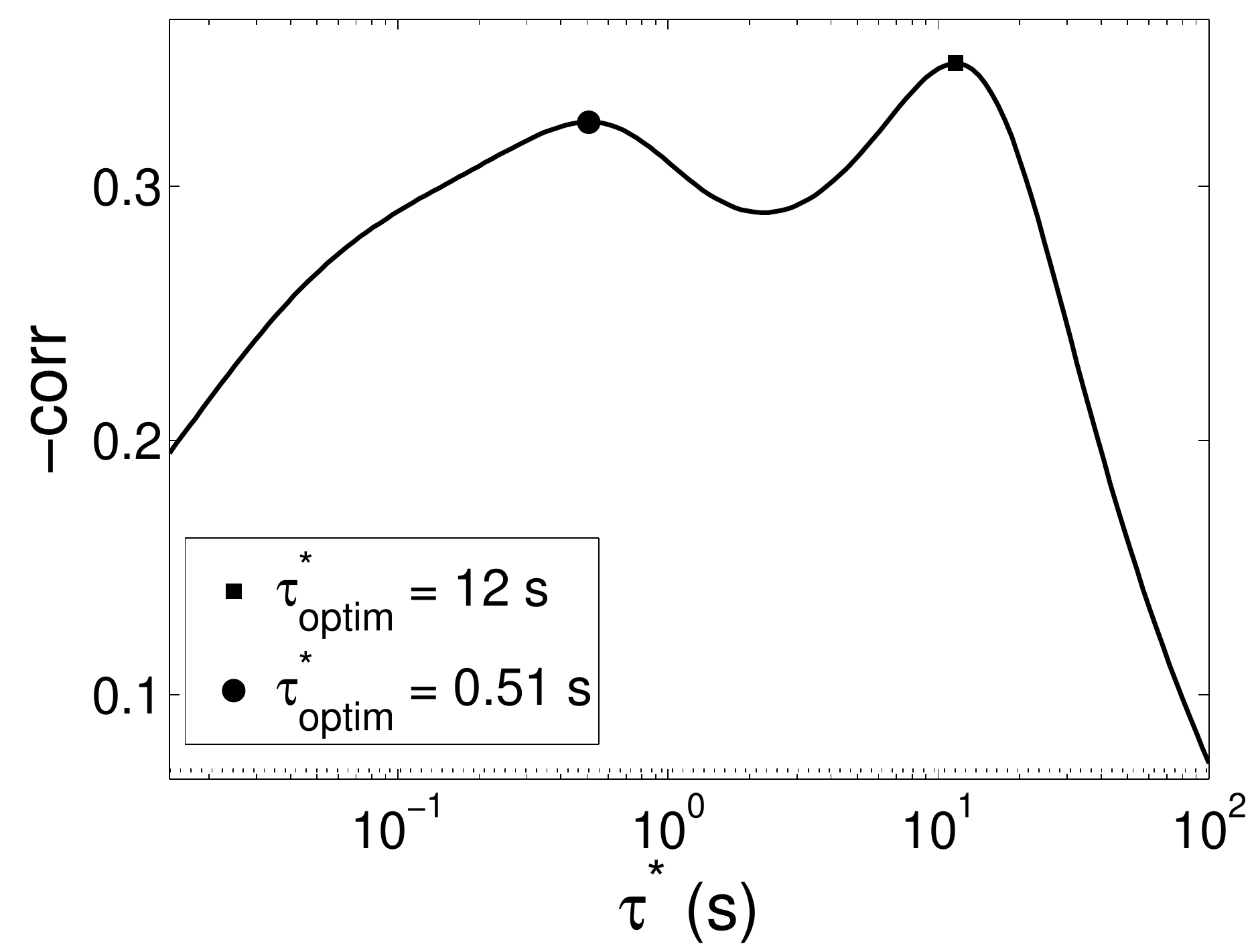,
    width=0.45\unitlength,
    }
  }
  \put(0.5,0)
  {
    \epsfig
    {
    file=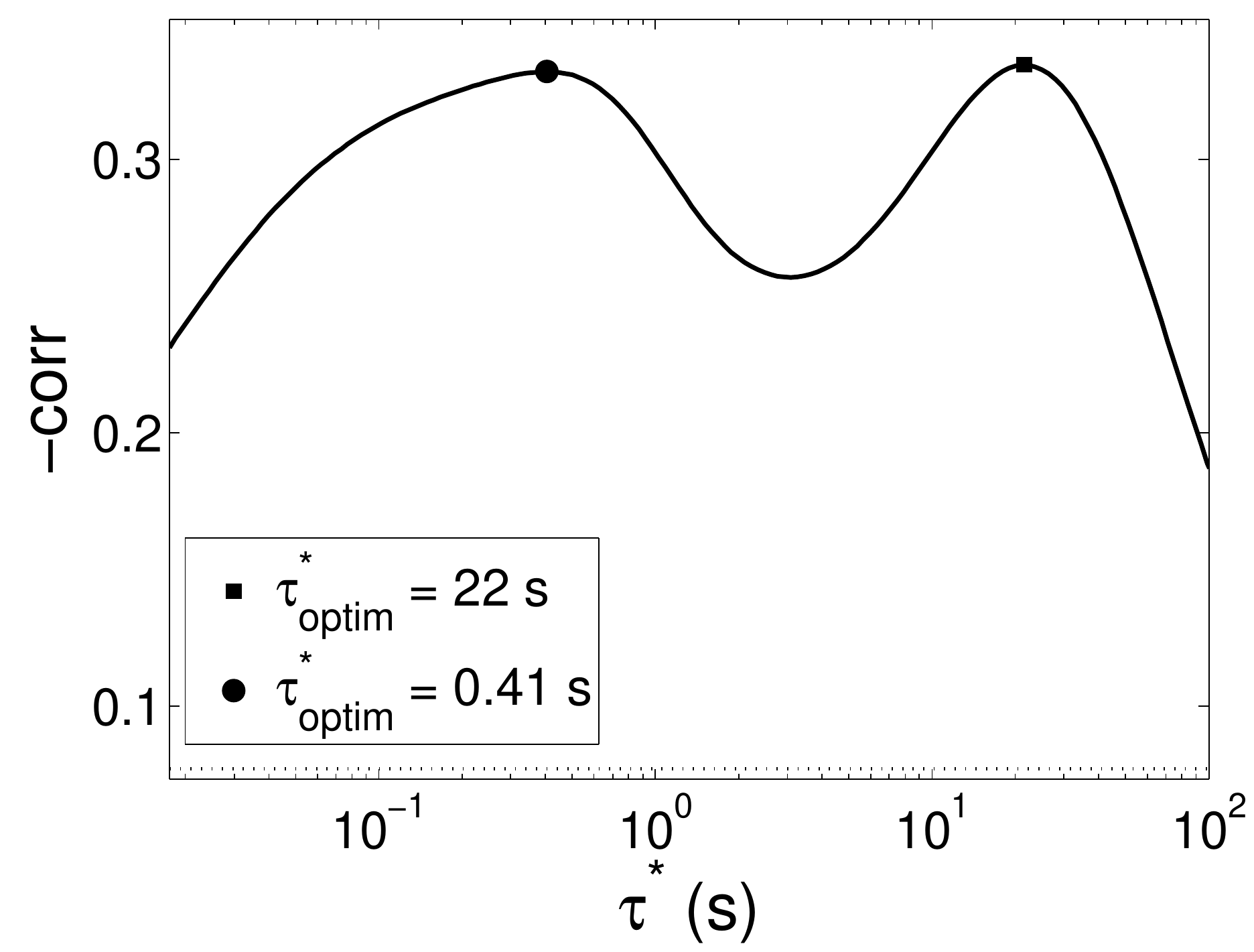,
    width=0.45\unitlength,
    }
  }
  \put(0.085,0.295){\makebox{\Large\textbf{A}}}
  \put(0.595,0.295){\makebox{\Large\textbf{B}}}
\end{picture}
\caption{{\bf Slow time-scales inference procedure on simulation data with \STD{} and spike-frequency adaptation.}
Correlation between low-pass filtered network activity $f$ (see Eq.~\ref{eq.linearIntegrator}) and the size of the immediately subsequent network spike plotted against the time-scale $\tau^*$ of the low-pass integrator. In this case, the mean-field model includes, besides short-term depression (\STD{}), a mechanism mimicking spike-frequency adaptation. Panel A: spike-frequency adaptation with characteristic time $\tau_{\mathrm{SFA}} = 15$ s. Panel B: $\tau_{\mathrm{SFA}} = 30$ s. In both cases the correlation presents a \STD{}-related peak at around $\tau^* \simeq 500$ ms ($\tau_{\mathrm{STD}} = 800$ ms), consistently with Fig.~\ref{figure7}. The peaks at higher $\tau^*$s, found respectively at $11$ and $18$ s, roughly preserve the ratio of the corresponding $\tau_{\mathrm{SFA}}$ values.}
\label{figure9}
\end{center}
\end{figure}

To test the plausibility of this interpretation, we simulated networks with simultaneously active \STD{} and spike-frequency adaptation (\SFA{}, see Models and Analysis). Fig.~\ref{figure9} shows the results of time scale inference for two cases sharing the same time scale for \STD{} (800 ms) and time scale of \SFA{} differing by a factor of 2 ($\tau_{\mathrm{SFA}} = 15$ and 30 s respectively). In both cases the negative correlation peaks at around $\tau^* \simeq 500$ ms; this peak is plausibly related to the characteristic time of \STD{}, consistently with Fig.~\ref{figure7}. The peaks at higher $\tau^*$s, found respectively at $12$ and $22$ s, roughly preserve the ratio of the corresponding $\tau_{\mathrm{SFA}}$ values.

This analysis provides preliminary support to the above interpretation of the double peak in Fig.~\ref{figure8}, right panel, in terms of two coexisting fatigue processes with different time scales.

We also checked to what extent the avalanche sizes were influenced by the immediately preceding amount of available synaptic resources $r$, and we found weak or no correlations; this further supports the interpretation offered at the end on the previous section, that avalanches are a genuine manifestation of the network excitability which amplifies a wide spectrum of small fluctuations.

\section*{Discussion}
Several works recently advocated a key role of specific network connectivity topologies in generating `critical' neural dynamics as manifested in power-law distributions of avalanches size and duration (see \cite{friedman2013hierarchical,massobrio2014criticality}). Also, it has been suggested that `leader neurons', or selected coalitions of neurons, play a pivotal role in the onset of network events (see e.g. \cite{eckmann2008leader,shein2008management,pirino2014topological,luccioli2014clique}). While a role of network topology, or heterogeneity in neurons' excitability, is all to be expected, we set out to investigate what repertoire of network events is accessible to a network with the simplest, randomly sparse, connectivity, over a wide range of excitation-inhibition balance, in the presence of \STD{} as an activity-dependent self-inhibition.
In the present work we showed that network spikes, avalanches and also large fluctuations we termed `quasi-orbits' coexist in such networks, with various relative weights and statistical features depending on the excitation-inhibition balance, which we explored extensively, including the role of finite-size noise (irregular synchronous regimes in balanced excitatory-inhibitory networks has been studied in \cite{benayoun2010avalanches}). We remark in passing that the occurrence of quasi-orbits is primarily related to the proximity to a Hopf bifurcation for the firing rate dynamics; on the other hand, the occurrence of \NS{} and, presumably, avalanches, does not necessarily require this condition: for instance, \NS{} can occur in the proximity of a saddle-node bifurcation, where the low-high-low activity transitions derive from the existence of two fixed points, the upper one getting destabilized by the fatigue mechanism (see e.g. \cite{gigante2007diverse,mejias2010irregular}); notably, in \cite{millman2010self} the authors find that, in a network of leaky integrate-and-fire neurons endowed with \STD, when a saddle-node separates an up- and a down-state, the dynamics develops avalanches during up-state intervals only.
We also remark that, with respect to the power-law distribution of avalanches, it is now widely recognized that while criticality implies power-law distributions, the converse is not true, which leaves open the problem of understanding what is actually in operation in the neural systems observed experimentally (for a general discussion on the issues involved, see \cite{beggs2012being}). In the present work, we do not commit ourselves to the issue of whether avalanches could be considered as evidence of Self-Organized Criticality.

In summary, the main contributions of the present work can be listed as follows.

We present a low-dimensional network model, derived from the mean field theory for interacting leaky integrate-and-fire neurons with short-term depression, in which we include the effect of finite-size (multiplicative) noise.

At the methodological level we introduce a probabilistic model for events detection, and a method for inferring the time-scale(s) of putative fatigue mechanisms. At the phenomenological level we recognize the existence of quasi-orbits as an additional type of network event, we show the coexistence of quasi-orbits, network spikes, and avalanches, and study their different mixing depending on the excitability of the network. We also offer a theoretical interpretation of the phenomenology, through a bifurcation analysis of the mean-field model, and a prediction on the effect of noise in the proximity of a Hopf bifurcation.

\section*{Acknowledgments}
We thank Neta Haroush and Maurizio Mattia for several stimulating discussions.

\nolinenumbers

\end{document}